\documentclass[12pt]{article}
\usepackage{a4wide,amsmath,epsfig}
\usepackage{longtable}
\allowdisplaybreaks[4] 
\newcommand{\ba}{\begin{array}{c}}
\newcommand{\ea}{\end{array}}
\newcommand{\nn}{\nonumber}

\newcommand{\wms}{\widetilde{MS}}
\newcommand{\be}{\begin{equation}}
\newcommand{\ee}{\end{equation}}
\newcommand{\chpt}{$\chi$PT}
\newcommand{\rcht}{R$\chi$T}
\newcommand{\ket}{\,\rangle}
\newcommand{\bra}{\langle \,}

\newcommand{\mL}{\mathcal{L}}

\newcommand{\mO}{\mathcal{O}}

\newcommand{\hM}{\hat{M}}
\newcommand{\hc}{\hat{c}}
\newcommand{\hd}{\hat{d}}
\newcommand{\Frac}[2]{\frac{\displaystyle #1}{\displaystyle #2}}
\newcommand{\cO}{{\cal O}}

\newcommand{\lsim}{\stackrel{<}{_\sim}}
\newcommand{\gsim}{\stackrel{>}{_\sim}}
\begin{document}

%%%%%%%%%%%%%%%%%%%%%%%%%%%%%%%%%%%%%%%%%%%%%%%%%%%%%%%%%%%%%%%%%%%%%%%%%%%%%%%

\begin{flushright}
 PUPT-2325
\end{flushright}

\rightline{\today}

\vspace*{1cm}

\begin{center}
{\Large \bf High energy constraints in the octet $SS-PP$ correlator
\\[0.3cm]
and resonance saturation at NLO in $\bf 1/N_C$}\\
[2 cm]

{\bf Juan Jose Sanz-Cillero$^{1}$ and Jaroslav Trnka$^{2,3}$}
\\[1.2 cm]
$\ ^{1}${\it Grup de F\`\i sica Te\`orica and IFAE, Universitat
Aut\'onoma de Barcelona,\\ E-08193 Barcelona, Spain}
,\\[0.4cm]
$\ ^{2}${\it Department of Physics, Princeton University, 08540 Princeton, NJ,
USA},\\[0.4cm]
$\ ^{3}${\it Institute of Particle and Nuclear Physics, Faculty of
Mathematics and Physics,\\ Charles University in Prague, 18000
Prague,
Czech Republic.}
\end{center}

\vspace*{2.0cm}

\begin{abstract}
We study the octet $SS-PP$ correlator within resonance
chiral theory
%%
%%, the chiral invariant  theory for Quantum Chromodynamics in the resonance region.
%%
%%In this work we provide the quantum field theory
%%calculation of this correlator
%%
up to the one-loop level, i.e., up to next-to-leading order
in the $1/N_C$ expansion.
%%
%%We will work within the single resonance approximation and
%%our  correlator will be  used
%%as an interpolator between low and high energies.
%%Thus,
%%
We will require that  our correlator follows
the power behaviour prescribed by the operator product expansion
at high euclidian momentum. Nevertheless,
we will not make use of short-distance constraints from   other
observables.  Likewise, the high-energy behaviour will be demanded
for the whole correlator, not for individual absorptive channels.
%%
%%as it was done in a previous resonance chiral theory  calculation.
%%
%%We find pretty bad low-energy constant predictions  when
%%only the simplest resonance lagrangian is considered,  with operators with
%%at most one resonance field.  This simple action can properly describe
%%just the $\pi\pi$ cut while the description of other two-meson channels
%%is inadequate ($S\pi$, $V\pi$, $P\pi$\dots).
%%One must add the relevant operators with two resonance fields
%%in order to  get the right momentum dependence for the $R\pi$ channels.
%%
The amplitude is progressively improved by considering more and more complicated
operators in the  hadronic lagrangian.
Matching the resonance chiral theory result with chiral
perturbation theory at low energies  produces the estimates
$L_8(\mu)^{SU(3)} = (1.0\pm 0.4)\cdot 10^{-3}$ and
$C_{38}(\mu)^{SU(3)} = (8\pm 5)\cdot 10^{-6}$ for $\mu=770$~MeV.
%%
%%These low-energy constants are in general found in agreement
%%with the values from a previous next-to-leading order analysis
%%based on dispersion relations and from other
%%former studies in the bibliography.
%%
The effect of  alternative renormalization schemes is also
discussed in the article.
\end{abstract}

\newpage

\setcounter{footnote}{0}

%%%%%%%%%%%%%%%%%%%%%%%%%%%%%%%%%%%%%%%%%%%%%%%%%%%%%%%%%%%%%%%%%%%%

\tableofcontents

\section{Introduction}

The effective field theory (EFT) approach is a very powerful tool
for the investigation of Quantum Chromodynamics (QCD) at long distances.
Chiral Perturbation theory ($\chi$PT)~\cite{Wein,Gasser1,Gasser2} is the EFT
for the description of the chiral (pseudo) Goldstones in the low energy domain
$p^2 \ll \Lambda_H^2\sim 1$~GeV$^2$, with $\Lambda_H$ typically the scale of
the lowest resonance masses. The calculation of the QCD matrix elements is
then organized at long distances
in growing powers of the external momenta and  light quark  masses.
Recent progress has allowed to carry $\chi$PT
up to ${\cal O}(p^6)$, i.e., up to the two-loop
level~\cite{op6-chpt,SS-PP-Bijnens,fit10,new-fit10}.

In the intermediate resonance region, $\Lambda_H \lsim E \lsim  2$~GeV,
$\chi$PT stops being valid and one must explicitly include
the resonance fields  in the Lagrangian description.
Unfortunately, this is not a straightforward process because there is no natural
expansion parameter in this region as several relevant mass scales appear
in this range (resonance masses, momenta, widths, the characteristic \chpt\ loop scale
$\Lambda_{\chi}\sim 4\pi F$...).
Resonance Chiral Theory (R$\chi$T)  describes the interaction
of resonance and pseudo-Goldstones within a general chiral invariant
framework~\cite{Ecker1, Ecker2}. Alternatively to the chiral counting,
it uses the $1/N_C$ expansion of QCD in the limit of large number of colours~\cite{NC}
as a guideline to organize the perturbative expansion.
At leading order (LO), just tree-level diagrams contribute
while loop diagrams yield higher order effects.
Integrating out the heavy resonance states leaves at low energies the
corresponding chiral invariant effective theory, $\chi$PT.
Many works have investigated various aspects of R$\chi$T: equivalence of
formalisms~\cite{Ecker2,Trnka1,Trnka2,Bijnens1};
Green functions~\cite{Pich1,rcht-op6,Pich3,Pich4,GF,Prades-GF,Knecht};
applications to
phenomenology~\cite{Pich1,Karel,Roig,Ivashyn1,Ivashyn2,Juanjo6,Jamin-cdcm,Masjuan-cd};
determination of chiral low-energy constants (LECs)
at NLO in $1/N_C$~\cite{Karel,L10-Cata,L9,L8,L10};
determination of the one-loop ultraviolet divergence structures in the
generating functional~\cite{Rosell-genera};
implications about the renormalizability~\cite{Juanjo3,Juanjo1};
possible issues with extra degrees of freedom
in the renormalized propagator~\cite{Trnka3,Trnka4};
renormalization group studies~\cite{RGE}.

The infinite tower of mesons contained in large--$N_C$ QCD is often truncated
to the lowest states in each channel, usually named as single
resonance approximation (SRA).
This approximation has led to successful predictions
of ${\cal O}(p^4)$ and  ${\cal O}(p^6)$ low-energy constants
(LECs)~\cite{Ecker1,Ecker2,Karel,Guo-aIJ,PI:08}.
However, the study of Regge models with an infinite number
of mesons has shown that if one keeps just the lightest states
with exactly the same couplings and masses of the full model then
one finds problems  in the short-distance matching and
wrong values are obtained for the LECs~\cite{Golterman-L8}.
Thus, in a high-energy matching with the operator
product expansion (OPE)~\cite{SVZ}
the parameters of the truncated theory will be shifted in order to accommodate
the right short-distance dependence.
Chiral symmetry ensures the proper low-momentum structure of the \rcht\ amplitudes
around $p^2=0$  but their high energy behaviour is not fixed by symmetry alone.
%%
%%Nevertheless, one knows that for large Euclidean momenta, $(-p^2)\gsim 2$~GeV$^2$
%%the $SS-PP$ correlator is expected to follow a vanishing behaviour prescribed
%%by  the OPE.
In that sense, the matched amplitude can be understood with the help
of Pad\'e approximants as an rational interpolator between the deep
Euclidean $p^2=-\infty$ and $p^2=0$~\cite{Masjuan-Pade,MHA}.   The
Weinberg sum-rules (WSR)~\cite{Weinberg2} yield the most convenient
parameters for the interpolation rather than the accurate
determinations of the resonance couplings. Furthermore, the \rcht\
couplings for the lightest mesons are expected to be in better
agreement, whereas the parameters from the highest excitations may
lie far from their right values~\cite{Masjuan-Pade}.

The connection of the \rcht\ amplitudes with the operator product
expansion~(OPE)  at high energies
seems {\it a priori} a useful procedure to include extra information from QCD
in the resonance theory.  It allows to fix combinations of couplings
(e.g., through WSR),  decreasing the number
of unknown parameters in the analysis.
%%
%%This must not be understood truly as the description of the amplitude at
%%infinite momentum but rather as a procedure
%%to incorporate the right energy dependence of the OPE in the
%%Euclidean range that overlaps
%%with the region where \rcht\ can still valid.
However, large--$N_C$ QCD has an infinite number of hadrons and
%%
%%and one should approach this limit to recover the precise result.
%%By adding more and more states to the theory (and eventually an infinite number
%%of them)  one expects to reproduce better the amplitude up to higher energies.
%%
in order to reproduce the full large--$N_C$ theory one must consider
the tree-level exchanges of heavier and heavier resonances. In the
hadronical ansatz approach, one adds more and more poles to the
rational approximant~\cite{Masjuan-Pade,MHA}. Equivalently, this can
be realized  within the quantum field theory  framework as
a generating functional with a lagrangian including interaction
operators $J - R_j$ that couple the external current source $J$ and
heavier and heavier resonances $R_j$ (e.g.   of the form $c_{m,j}
\bra S_j \chi_+\ket$ for the $SS$ correlator).

The extension of R$\chi$T beyond the tree level approximation  still needs
to be worked out in detail. Although some theoretical issues
on the renormalizability of \rcht\
still need further clarification~\cite{Juanjo3,Juanjo1,prop},
several chiral LECs have been already computed up to NLO in $1/N_C$
through quantum field theory (QFT) one-loop calculations~\cite{L10-Cata,L9}
and dispersion relations~\cite{L8,L10}.
In this article we will focus our attention on the
chiral octet $SS-PP$ correlator (for instance, with $I=1$),
which in the chiral limit   is
determined at low energies by the $\cO(p^4) $ and $\cO(p^6)$ LECs, respectively,
by $L_8$~\cite{Gasser2} and $C_{38}$~\cite{op6-chpt}.
The  correlator is computed up to
next-to-leading order in $1/N_C$ (NLO) and the chiral limit will be
assumed all along the article.

At the one-loop level --NLO in $1/N_C$--,   one needs also to devise a procedure
to reach the infinite resonance limit of large--$N_C$ QCD.
In the case of two--point Green-functions, the imaginary part of the
one-loop diagrams is given through the optical theorem
by the square of two-meson form-factors computed at tree-level.
Thus, based on a dispersive approach,
one may add the contribution to the spectral function from
higher and higher two-meson absorptive cuts
by providing the corresponding form-factors~\cite{L8,L10}.
This would be, in some sense, the natural extension of
the minimal hadronical ansatz~\cite{MHA} to the one-loop situation.
In a previous computation of the octet $SS-PP$ correlator up to NLO in $1/N_C$,
the intermediate two-meson channels
were analyzed individually~\cite{L8}.
The corresponding tree-level form-factors were made to vanish appropriately at high
energies~\cite{L10,Rosell-thesis}.    This allowed to recover the
correlator from its spectral function through an unsubtracted dispersion relation.
However, in general, it is not always possible  to fulfill
the high-energy constraints for all the form-factors at
once~\footnote{In the case of the scalar and pseudo-scalar form-factors,
it is still possible to impose the right high-energy behaviour to all the
form-factors if one considers operators with two and three resonance fields
$\mL_{RR'}$ and $\mL_{RR'R''}$~\cite{L10,Rosell-thesis}.
Nonetheless, there is no consistent set of constraints  for
all the vector and axial-vector form-factors if only a finite number
of resonances is considered~\cite{L10,Rosell-thesis}.
A similar kind of  inconsistences  was found in the study of
three--point Green-functions at large $N_C$~\cite{Prades-GF}.}.
Only the two-meson absorptive cuts with at most one resonance ($\pi\pi$ and $R\pi$)
were considered in Ref.~\cite{L8},
as the $RR'$ channels have their thresholds at $(M_R+M_{R'})\sim
2$~GeV and are suppressed at low energies.
Likewise, the short-distance constraints from $VV-AA$ Weinberg
sum-rules and the $\pi\pi$ vector and the scalar form-factor were
used there in order to fix some of the couplings appearing in the
analysis.

In the quantum field theory approach proposed in this work, one  has
a mesonic lagrangian   which at the classical level
generates the large--$N_C$ amplitudes and whose quadratic fluctuations
around the classical field configuration provide the
one-loop corrections~\cite{Rosell-genera}. The complete QCD
generating functional is approached as one adds more and more
hadronic operators to the action. Eventually, one should add the
infinite number of possible terms of the given $1/N_C$ order under
consideration. For instance, the $S\pi\pi$
interaction (provided by $c_d\bra S u_\mu u^\mu\ket$~\cite{Ecker1})
is of the same order as in $1/N_C$ as the $SP\pi$ vertex
(given by the $\lambda_1^{SP}\bra \{ \nabla^\mu S, P\}u_\mu\ket$
operator~\cite{rcht-op6,L10,Rosell-thesis}).
Notice that one never has a complete
description with a finite number of operators. The basic lagrangian
$\mL_G+\mL_R$ with at most one resonance field in each term~\cite{Ecker1}
provides an incomplete description of the $R\pi$ channels, as the possible
diagrams with $R'$ resonances exchanged in the $s$--channel are
missing~\cite{L8,L10}.   This requires the incorporation of
operators $\mL_{RR'}$ with two resonance
fields~\cite{rcht-op6,L10,Rosell-thesis}. In the same way, the $RR'$
absorptive cuts are now badly described without the $\mL_{RR'R''}$ terms
with three resonance fields.

The chiral structure of the lagrangian ensures the right structure
at long distances. On the other hand,   we will impose that the
correlator follows the short-distance behaviour prescribed by the
OPE. The one-loop \rcht\ amplitude will be used as an improved
interpolator between low and high energies. The resonance couplings
become then  interpolating parameters that must approach their
actual values  in the full QCD as more and more operators are added
to the \rcht\ action. On the contrary to  what was done in former
works~\cite{L8,L10}, the short-distance matching
will  be carried out in the present article for the total correlator
and  spectral function~\cite{NLO-satura}, rather than for
individual channels. Likewise, we will not use the short-distance
constraints from other amplitudes to fix the couplings in the
one-loop correlator.
We will work within the SRA, including just the chiral Goldstones and
the lightest multiplets of scalar, pseudo-scalar, vector and axial-vector resonances.
In a first step, the $SS-PP$ correlator will be computed
at NLO in $1/N_C$ with the simplest \rcht\
lagrangian, with operators with at most one resonance field
($G_V$, $c_m$, $d_m$...)~\cite{Ecker1}.
This provides the proper
structure for the intermediate tree-level exchanges
($\pi,\, S,\, P$ one-particle channels)
and the two-Goldstone cut $\pi\pi$.  However, this simple lagrangian
fails to describe the
$R\pi$ and $RR'$ channels as the lagrangian~\cite{Ecker1}  makes their
form-factors behave like a constant or like a growing power of the momentum
at high energies~\cite{L9,L8,L10,Rosell-thesis,Cillero-thesis}.
This will be partly cured by the consideration of
$\lambda^{RR'}$ operators with two resonance
fields~\cite{rcht-op6,L10,Rosell-thesis,Cillero-thesis},
which now allow an appropriate description of the $R\pi$ channels,
though the $RR'$ ones still behave badly.
Although these cuts with two resonances were neglected in the
dispersive approach~\cite{L8},
removing part of the one-loop diagrams is not theoretically well defined
and may lead to inconsistences in the renormalization of the QFT.
Furthermore,  it is not trivial that the effect of the
$RR'$ cuts in the short-distance matching is fully negligible.
Hence, all the possible diagrams contributing to the correlator
up to NLO will be kept in our study.

The amplitude is first computed within the usual subtraction scheme of $\chi$PT~\cite{Gasser1} (denoted for simplicity as $\wms$
all along the article).
However, though equivalent at low
energies, some appropriate schemes will be  found  more convenient:
pole masses and other schemes that minimize the uncertainties
derived from the short-distance constraints. This will help us to
determine the $\cO(p^4)$ and $\cO(p^6)$  LECs, respectively
$L_8(\mu)$ and $C_{38}(\mu)$. The high-energy constraints  and their
meaning  will be discussed and the convergence to full large--$N_C$
QCD will be tested as  more and more hadronic operators are added to
the \rcht\  action. This work is thought as a complementary and an
alternative approach to the dispersive analysis in Ref.~\cite{L8}.

The article is organized as follows. Resonance chiral theory  is
introduced in detail in Sec.~\ref{sec.rcht}.  The
octet $SS-PP$ correlator is defined
in Sec.~\ref{sec.SS-PP} and its one-loop \rcht\ computation is provided
in Sec.~\ref{sec.one-loop}.  The high-energy constraints and low energy expansions
are respectively given in Secs.~\ref{sec.high} and~\ref{sec.low}.
The contributions from operators $\mL_{RR'}$ with two resonance fields
have been singled out in Sec.~\ref{sec.extra-operators}
to ease the main argumentation of the article.   Finally,
the phenomenological analysis is given in Sec.~\ref{sec.pheno} and
the conclusions are provided in Sec.~\ref{sec.conclu}.
Some technical results  are relegated to the Appendices.

\section{Resonance chiral theory lagrangian}
\label{sec.rcht}

Within the large--$N_C$ approach the mesons will be classified
within $U(3)$ multiplets.  The chiral Goldstone bosons are introduced
by means of the basic building block,
\begin{equation}
  u(\phi) = \exp\left(i\frac{\phi}{\sqrt2 F}\right)
\end{equation}
where $\phi=\frac{1}{\sqrt{2}}\lambda^a\phi^a $ and
\begin{equation}
\phi(x)\,\, = \,\,\left(
   \begin{array}{ccc}
   \frac{1}{\sqrt{2}}\pi^0 +\frac{1}{\sqrt6}\eta_8 +\frac{1}{\sqrt3}\eta_1
   &  \pi^+ &  K^+ \\
    \pi^- & -\frac{1}{\sqrt2 }\pi^0+\frac{1}{\sqrt6}\eta_8 +\frac{1}{\sqrt3 }\eta_1&
     K^0 \\
    K^- &  \overline{K}^0 & -\frac{2}{\sqrt6}\eta_8+\frac{1}{\sqrt3 }\eta_1 \\
   \end{array}
\right)\, .
\end{equation}
This forms the basic covariant tensors,
\begin{eqnarray}
 u_\mu &=&  i \, \{ u^\dagger (\partial_\mu - i r_\mu) u  \,
-\, u \, (\partial_\mu- i \ell_\mu) u^\dagger \}\, ,
\nn
\\
\chi_\pm &=& u^\dagger \, \chi\, u^\dagger \, \pm \, u\, \chi^\dagger \, u \, ,
\label{eq.bricks}
\\
 f_\pm^{\mu\nu} &=& u\, F_L^{\mu\nu}\, u^\dagger \,
 \pm \, u^\dagger \, F_R^{\mu\nu}\, u\, ,
 \nn
\end{eqnarray}
with $\chi= 2B_0(s+i p)$ containing the scalar and pseudo-scalar
external sources, $s$ and  $p$ respectively,  the right and left
sources $r^\mu$ and $\ell^\mu$
providing the vector and axial-vector external sources,
$v^\mu=\frac{1}{2}(r^\mu+\ell^\mu)$ and
$a^\mu=\frac{1}{2}(r^\mu -\ell^\mu)$ respectively,
and $F_{L,R}^{\mu\nu}$  the corresponding left and right field-strength tensors.

The Goldstone bosons are parametrized by the elements $u(\phi)$ of
the coset space $U(3)_L\times U(3)_R/U(3)_V$, transforming as
\begin{equation}
  u(\phi) \mapsto V_R u(\phi) h(g,\phi)^{-1} = h(g,\phi)u(\phi)V_R
\end{equation}
under a general chiral rotation $g=(V_L,V_R)\subset G$ in terms of
the $U(3)_V$ compensator field $h(g,\phi)$. This makes the tensors
$X=u^\mu,\chi_\pm ,f_\pm^{\mu\nu}$ to transform covariantly in the form,
\begin{equation}
X\,\mapsto\,  h(g,\phi)\, X\,  h(g,\phi)^{-1}\, .
\label{eq.transform}
\end{equation}

\subsection{Leading order lagrangian}

For the classification of the vertices entering in the
tree-level and one-loop amplitudes it will be useful to organize
the operators  of the  \rcht\ lagrangian according to the number of
resonance fields:
\begin{equation}
\mL\, \,=\, \,\mL_G\,\, +\,\mL_R\,\, +\,\, \mL_{RR'}\,\,+\,\,\ ...
\end{equation}
where $\mL_G$ only contains Goldstone bosons and external sources, $\mL_R$
also includes one resonance, etc.
Although in principle one should consider all the terms
compatible with symmetry,  most of the large--$N_C$
phenomenological calculations
consider operators with the minimal number of derivatives~\cite{PI:08}.
This is usually justified through the argument that
higher derivative operators tend to violate the asymptotic high energy QCD
behaviour~\cite{Ecker2,PI:08}. Likewise, its has been proven
in several cases that higher derivative resonance operators can be
removed from the hadronic action through meson field redefinitions
in the generating
functional~\cite{L9,Rosell-genera,Juanjo3,Juanjo1,Rosell-thesis,Cillero-thesis}.
In the present article, the leading lagrangian will only contain
operators at most $\cO(p^2)$, with the external sources
counted as $v^\mu, a^\mu\sim \cO(p)$ and
$\chi\sim \cO(p^2)$~\cite{Rosell-thesis,NLO-satura}.

The Lagrangian with only Goldstones has the same form as in
$\chi$PT but the coupling constants are different. In $\chi$PT we
have the leading order Lagrangian
\begin{equation}
{\cal L}_{\rm \chi PT}^{(2)} = \frac{F^2}{4}\langle u_\mu u^\mu + \chi_+
\rangle\, .
\end{equation}
In R$\chi$T beyond leading order
the constants standing in front of the operators
$\bra u^\mu u_\mu\ket$ and $\bra \chi_+\ket$
may not be the same as in $\chi$PT. Therefore,
generally we can write
\begin{equation}
{\cal L}_{G}  = \frac{\widetilde{F}^2}{4}\langle u^\mu
u_\mu \rangle + \frac{\hat{F}^2}{4}\langle \chi_+ \rangle
\end{equation}
where we explicitly distinguish between $\widetilde{F}$ and
$\hat{F}$. These can be split in the way,
\begin{equation}
\widetilde{F} = F + \delta \widetilde{F}, \qquad \hat{F} = F + \delta
\hat{F}
\end{equation}
where at large $N_C$ one has the matching condition
$\widetilde{F} =\hat{F} =F$ and, hence,
$\delta\widetilde{F}$ and $\delta\hat{F}$ are NLO in $1/N_C$.
On the contrary to what happens in $\chi$PT, where the parameters
($F$ and $B_0$) which characterize the terms $\bra u^\mu u_\mu\ket$
and $\bra \chi_+\ket$ do not become renormalized,
in \rcht\ the couplings   of these two operators
are needed to make the physical amplitude finite.
For simplicity, we choose to keep the definitions
of the chiral tensors unchanged and to renormalize instead
$\widetilde{F}$ and $\hat{F}$,
as it was done in Refs.~\cite{Rosell-genera,Rosell-thesis}
with the notation $\alpha_1=\widetilde{F}^2/4$ and $\alpha_2=\hat{F}^2/4$.

The Goldstone bosons
couple to massive $U(3)$ multiplets of the type $V(1^{--})$,
$A(1^{++})$, $S(0^{++})$ and $P(0^{-+})$.  The vector multiplet, for instance,
is given by
\begin{equation}
V_{\mu\nu}\,\, = \,\,\left(
   \begin{array}{ccc}
   \frac{1}{\sqrt{2}}\rho^0 +\frac{1}{\sqrt6}\omega_8 +\frac{1}{\sqrt3}\omega_1
   &  \rho^+ &  K^{*\, +} \\
    \rho^- & -\frac{1}{\sqrt2 }\rho^0+\frac{1}{\sqrt6}\omega_8 +\frac{1}{\sqrt3 }
    \omega_1&
     K^{*\, 0} \\
    K^{*\, -} &  \overline{K}^{\, *\, 0} & -\frac{2}{\sqrt6}\omega_8+\frac{1}{\sqrt3 }
    \omega_1 \\
   \end{array}
\right)_{\mu\nu}\, ,
\end{equation}
where we use the antisymmetric tensor formalism for spin--1 fields
to describe the vector and axial-vector
resonances~\cite{Ecker1,Ecker2,Bijnens1}.

The resonance fields $R$  are chosen to
transform covariantly under the chiral group as in
Eq.~(\ref{eq.transform})~\cite{Ecker1}.
The free-field  kinetic term is given by
the operators
\begin{equation}
{\cal L}_{RR}^{\rm Kin}\,\,
=\,\, -\frac12 \langle \nabla^\mu R_{\mu\nu} \nabla_\alpha
R^{\alpha\nu}\rangle + \frac14 M_R^2 \langle
R_{\mu\nu}R^{\mu\nu}\rangle + \frac12 \langle \nabla^\alpha R'
\nabla_\alpha R'\rangle -\frac12 M_{R'}^2\langle R' R' \rangle \, .
\end{equation}\label{Lagr1}
where $R=V,A$ are vector and axial vector resonances and $R'=S,P$
are scalar and pseudoscalar resonances.
%%
%%The simplest interaction
%%Lagrangian that contributes to ${\cal O}(p^4)$ LECs is
%%

The interaction terms which are linear in the resonance fields can be
obtained from the seminal work~\cite{Ecker1}:
\begin{eqnarray}
{\cal L}_{R} &=&
c_d\langle S u^\mu u_\mu \rangle + c_m \langle S
\chi_+\rangle + id_m \langle P\chi_-\rangle
+\frac{F_V}{2\sqrt2}\langle V_{\mu\nu}
f_+^{\mu\nu}\rangle + \frac{iG_V}{2\sqrt2}\langle V_{\mu\nu}
[u^\mu,u^\nu]\rangle +\frac{F_A}{2\sqrt2}\langle A_{\mu\nu}
f_-^{\mu\nu}\rangle.
\nonumber\\
%%&&\hspace{1cm}+ .
\label{Lagr2}
\end{eqnarray}

For our analysis of the $SS-PP$ correlator, the relevant bilinear terms
will be~\cite{rcht-op6,Rosell-thesis,Cillero-thesis}
\begin{equation}
\mL_{RR'}\,=\, i \lambda_1^{PV} \bra [\nabla^\mu P,V_{\mu\nu}]\, u^\nu\ket
\,+\,   \lambda_1^{SA} \bra \{\nabla^\mu S,A_{\mu\nu}\}\, u^\nu\ket
\,+\,   \lambda_1^{SP} \bra \{ \nabla^\mu S,P\} \, u_\mu\ket\, .
\end{equation}

Only single flavor--trace operators are considered for the construction of
the large--$N_C$ lagrangian.
At tree-level, the octet $SS-PP$ correlator only gets contributions from
this kind of terms, even at subleading orders in $1/N_C$.
Operators with two or more traces might
appear in the vertices of one loop diagrams but, since these multi-trace
terms are $1/N_C$--suppressed, these contributions would go to
next-to-next-to-leading order and they will be neglected in the present work.

The previous operators provide an appropriate description of the
form factors with two Goldstones or one resonance and one Goldstone
in the final state.  We will perform our most elaborate analysis with
the lagrangian $\mL_G+\mL_{R}+\mL_{RR'}$,  with
at most two resonance fields.  As we will see in next sections,
the \rcht\ description will progressively approach
the actual QCD amplitude as more and more complicated operators are added.
However, although we expect the contributions from the
operators with three resonance fields
to the LECs to be negligible at our level of accuracy,
a further refinement is eventually possible by considering
these operators $\mL_{RR'R''}$ .

\subsection{Subleading Lagrangian}

At the loop level, one needs to
introduce new subleading operators in order to cancel the ultraviolet
divergences,  to renormalize R$\chi$T  and to make the amplitudes finite.
As the leading order lagrangian operators are $\cO(p^2)$,
the naive dimensional analysis tells us that at one loop one expects to find
$\cO(p^4)$ ultraviolet divergences, requiring the introduction
of NLO counter-terms with a higher number of derivatives.

The new operators with just Goldstone bosons required at NLO are,
for the $SS-PP$ correlator under consideration,
\begin{eqnarray}
{\cal L}_{GB}^{NLO} &=&
%%
%%\widetilde{L}_6\langle \chi_+\rangle^2
%%+ \widetilde{L}_7 \langle \chi_-\rangle^2 +
%%
\frac{\widetilde{L}_8}{2}\langle
\chi_-^2 + \chi_+^2\rangle +i\widetilde{L}_{11}\langle \chi_-(\nabla_\mu
u^\mu -\frac{i}{2}\chi_-)\rangle
%%\nonumber\\&&
-\widetilde{L}_{12}\langle (\nabla_\mu
u^\mu -\frac{i}{2}\chi_-)^2\rangle
+\Frac{\widetilde{H}_2}{4}\langle \chi_+^2-\chi_-^2\rangle\, .
\nonumber\\&&
%%
%%\nonumber\\
%%
%%
%%
%%\\
%%{\cal L}_{\widetilde{\chi}}^{(6)} &=& \widetilde{C}_{38}\langle
%%\chi_{+\mu}\chi_+^\mu\rangle \,.
%%
\end{eqnarray}
Though we use the same structure of terms as in $\chi$PT,
the \rcht\ couplings $\widetilde{L}_i$ are not the same as the chiral LECs $L_i$.
The $\widetilde{L}_i$ will contribute at low energies
to ${\cal O}(p^4)$ chiral couplings $L_i$. The latter
are dominantly saturated by resonances exchanges,
so $\widetilde{L}_i$ are considered to be suppressed and subleading in the
$1/N_C$ expansion.

In order to make the resonance propagator finite, one needs to
renormalize the mass and wave functions ($M_R^{(B)\,\,
2}=M_R^{r\,\,2}+\delta M_R^2$, $R^{(B)}=Z_R^{\frac{1}{2}} R^r$) and
to introduce at NLO in $1/N_C$ the kinetic operator
\begin{equation}
{\cal L}_{\rm Kin}^{NLO}\,  \, = \,\,
\Frac{X_R}{2} \langle R \nabla^4 R\rangle\, ,
\end{equation}
with  $R=S,P$. No terms with vector or axial-vectors are needed
for the  present NLO analysis of the $SS-PP$ correlator.

Likewise, the renormalization of the vertex functions $s(x)\to S$ and $p(x)\to P$
at NLO in $1/N_C$ will require of the linear terms,
\begin{eqnarray}
{\cal L}_R^{NLO} &=& \lambda^S_{18} \langle S\nabla^2
\chi_+\rangle \,\, +\,\, i\lambda^P_{13}\langle P\nabla^2
\chi_-\rangle \,.
\end{eqnarray}

At NLO in $1/N_C$, all these subleading counter-terms
can only contribute through tree-level diagrams.

\subsection{Equations of motion and redundant operators}

%%
%%If one considers just the interactions linear in the resonance fields,
%%
The equations of motion (EOM)  of the leading lagrangian
are given by~\cite{Rosell-thesis,Rosell-genera},
\begin{eqnarray}
\nabla^\mu u_\mu &=& \Frac{i}{2}\chi_-
+\Frac{i c_m}{F^2} \{ \chi_-,S \} -\Frac{d_m}{F^2} \{\chi_+,P\}
%%%\nonumber\\
%%%&&\qquad\qquad
%%%  -  \Frac{2 c_d}{F^2} \nabla^\mu\{u_\mu,S\}
%%%  -\Frac{1}{2 F^2}[u_\mu,[\nabla^\mu S,S]]
%%%   -\Frac{1}{2 F^2}[u_\mu,[\nabla^\mu P,P]]
%
%%%%\,\,\,+\,\,\, \cO(\phi\, R)\,\,\,+\,\,\, \cO(R^2)
%%%%\,+\, \cO(V)+\cO(A)\, ,
%
\,\,\, +\,\, \, ...
\label{eq.EOM1}\\
\nabla^2 S &=& -M_S^2 S + c_d u_\mu u^\mu + c_m \chi_+ \,\,\, +\,\,\,
%
%%%%\cO(R??????) \,,
%
...  \label{eq.EOM2}\\
\nabla^2 P &=& -M_P^2 P + i d_m \chi_-\,\,\, +\,\,\,
%%%%\cO(R??????) \,,
...    \label{eq.EOM3}
\end{eqnarray}
where the dots stand for terms with vector or axial-vector resonances or sources,
two-meson fields or
with one scalar-pseudoscalar external source  and one meson field.

Since most of the subleading resonance operators are proportional
to the EOM, it is possible to simplify our new NLO resonance
operators by means of appropriate meson field redefinitions,:
\begin{eqnarray}
\mL_{\rm Kin}^{NLO}\longrightarrow {\cal L}^{NLO,\,\,eff}_{{\rm Kin}}
 &=& -\lambda^S_{18}M_S^2\langle SS\rangle
+c_m\lambda^S_{18}\langle \chi_+^2\rangle
%%%\nonumber\\
%%%&&
-i\lambda^P_{13}M_P^2\langle PP\rangle -d_m\lambda^P_{13}\langle \chi_-^2\rangle
\,\,\, +\,\,\, ...
\nonumber\\
\nonumber\\
\mL_R^{NLO}\longrightarrow   {\cal L}_R^{NLO,\,\,eff}   &=&
\frac{X_SM_S^4}{2} \langle SS\rangle
+\frac{c_m^2X_S}{2} \langle \chi_+^2\rangle
%%
%%-c_dX_SM_S^2\langle  S u^\mu u_\mu\rangle
%%
- c_mX_SM_S^2 \langle S\chi_+\rangle\nonumber \\
 &&+  \frac{X_PM_P^4}{2}\langle PP\rangle -
\frac{d_m^2X_P}{2}\langle \chi_-^2\rangle - id_mX_PM_P^2\langle
P\chi_-\rangle\,\,\,+\,\,\, ...
\end{eqnarray}
where the dots stand for operators that do not contribute to the $SS-PP$
correlator at NLO.
After the field redefinition the resonance operators
$\mL_{\rm Kin}^{NLO}$ and $\mL_R^{NLO}$
disappear and the surviving terms in the \rcht\ lagrangian
carry in front the  effective combinations,
\begin{eqnarray}
\widetilde{L}_8^{eff} &=& \widetilde{L}_8 +\frac{1}{2}c_m^2X_S-\frac{1}{2}d_m^2X_P +
c_m\lambda^S_{18} - d_m\lambda^P_{13},\nonumber\\
\widetilde{H}_2^{eff} &=& \widetilde{H}_2 +c_m^2X_S+d_m^2X_P +
2c_m\lambda^S_{18} + 2d_m\lambda^P_{13},\nonumber\\
(M_S^2)^{eff} &=& M_S^2 - X_SM_S^4,\nonumber\\
(M_P^2)^{eff} &=& M_P^2 - X_PM_P^4,\nonumber\\
%%%
%%%c_d^{eff} &=& c_d - c_dX_SM_S^2\nonumber,\\
%%%
c_m^{eff} &=& c_m - c_mX_SM_S^2 -M_S^2\lambda^S_{18},\nonumber\\
d_m^{eff} &=& d_m - d_mX_PM_P^2 - M_P^2\lambda^P_{13}.
\label{eq.eff-couplings}
\end{eqnarray}
The $\widetilde{L}_{11}$ and $\widetilde{L}_{12}$ operators do not
contribute to terms which can be relevant to our amplitude up to NLO
and we will see that they are not present in the final result.

\section{Chiral octet $SS-PP$ correlator}
\label{sec.SS-PP}

In the case of $SU(3)$--octet quark bilinears,
the two-point Green function $SS-PP$ is defined as
\begin{equation}
\Pi_{S-P}^{ab}(p) \,\,=\,\, i\,\int d^4x e^{ip\cdot x}
\langle 0| T[S^a(x)S^b(0)-P^a(x)P^b(0)]|0\rangle
= \delta^{ab}\Pi(p^2)\,,
\end{equation}
with $S^a=\bar{q}\frac{\lambda_a}{\sqrt{2}} q$ and
$P^a=i\bar{q}\frac{\lambda_a}{\sqrt{2}}\gamma_5 q$,
being $\lambda_a$ the Gellmann matrices ($a=1,\dots 8$).

In the chiral limit, assumed all along the article, the low-energy
expansion of the octet correlators is determined by $\chi$PT
in the form~\cite{SS-PP-Bijnens},
\begin{eqnarray}
\Pi(p^2)_{\chi PT} &=& B_0^2\Bigg\{\frac{2F^2}{p^2}
+ \left[ 32L^r_8(\mu_\chi)
+\frac{\Gamma_8}{\pi^2}\left(1-\ln\frac{-p^2}{\mu_\chi^2}\right)
\right]
\label{ChPT}
\\&&\hspace{1cm}
+\frac{p^2}{F^2}\left[32C_{38}^r(\mu_\chi)
-\frac{\Gamma^{(L)}_{38}}{\pi^2}\left(1-\ln\frac{-p^2}{\mu_\chi^2}\right)
+ {\cal O}(N_C^0)\right] + {\cal O}(p^4)\Bigg\}\nonumber
\end{eqnarray}
where in $\Gamma_8=5/48$ [$3/16$] and $\Gamma^{L}_{38} = -5L_5/6$ [$-3 L_5/2$]
in $SU(3)$--$\chi$PT  [$U(3)$--$\chi$PT].
Notice that in $\chi$PT the correlator is exactly independent of
the renormalization scale $\mu_\chi$, being its choice completely arbitrary.

In the resonance region, one obtains at leading order in $1/N_C$,
\begin{equation}
\Pi (p^2)_{LO} \, =\,  \frac{2B_0^2F^2}{p^2} \, +\,  16B_0^2 \sum_{i}\left(
\frac{c_{m,i}^2}{M_{S,i}^2-p^2}\, -\, \frac{d_{m,i}^2}{ M_{P,i}^2-p^2}\right)\, ,
\label{LO}
\end{equation}
where one sums   over the different  resonance multiplets.  The subscript $_{,i}$ in
$M_{R,i}$, $c_{m,i}$ and $d_{m,i}$ refers to the coupling of the $i$--th
resonance multiplet of the corresponding kind.
The requirement of
the high energy OPE behaviour
$\Pi(p^2)\stackrel{p^2\rightarrow\infty}{\sim} 1/p^6$
produces the short-distance conditions
~\footnote{The
tiny dimension four condensate
$\frac{1}{B_0^2}\, \langle \cO_{(4)}^{^{SS-PP}}\rangle\simeq
- 12 \pi\alpha_S F^4$ will be neglected in this work~\cite{PI:08,Peris-O4}.
}~\cite{PI:08}
\begin{equation}
\sum_{i}(c_{m,i}^2-d_{m,i}^2) = \frac{F^2}{8},\qquad\qquad
\sum_{i}c_{m,i}^2M_{S,i}^2-d_{m,i}^2M_{P,i}^2 = 0\, .
\label{eq.WSR}
\end{equation}
In the single resonance approximation (SRA),
it is then possible to express $c_m$ and $d_m$ in terms of $F$ and resonance masses,
\begin{equation}
c_m^2 \,\, =\,\, \frac{F^2}{8}\,\frac{M_P^2}{M_P^2-M_S^2}\qquad\qquad\qquad
d_m^2 \,\,=\,\, \frac{F^2}{8}\,\frac{M_S^2}{M_P^2-M_S^2}.
\label{eq.WSR-solution}
\end{equation}
At low energies, we can match the large--$N_C$ expression~(\ref{LO}) with
the \chpt\ expression~(\ref{ChPT}),  obtaining
the LO prediction for the low energy coupling constants $L_8$ and
$C_{38}$,
\begin{eqnarray}
L_8 &=& \frac{c_m^2}{2M_S^2} - \frac{d_m^2}{2M_P^2} =
\frac{F^2}{16}\left(\frac{1}{M_P^2}+\frac{1}{M_S^2}\right),\\
C_{38} &=& \frac{c_m^2F^2}{2M_S^4} -\frac{d_m^2F^2}{2M_P^4} =
\frac{F^4}{16M_P^2M_S^2}\left(1+\frac{M_P^2}{M_S^2}+\frac{M_S^2}{M_P^2}\right)
\end{eqnarray}
%%
%%The numerical value strongly depends on input parameters.
%%
For the inputs $M_S=M_P/\sqrt2\simeq 1$~GeV, one obtains $L_8\approx
0.7\cdot 10^{-3}$, $C_{38} \approx 7\cdot 10^{-6}$ for $M_S=1\,{\rm
GeV}$.
However, one does not know to what renormalization scale $\mu_\chi$
these numerical predictions correspond.
In order to pin down this $\mu$--dependence,
one must carry the calculation up to the loop level.

\section{One-loop computation in resonance chiral theory}
\label{sec.one-loop}

We follow the renormalization procedure presented in \cite{L9}.
In general,
we will use   dimensional regularization and the ${\overline{MS}}-1$
subtraction scheme, usually employed  in $\chi$PT
calculations~\cite{Gasser1,Gasser2}.
This means  we will absorb
in the coupling counter-terms  the ultraviolet divergent piece from the loops,
counter-terms,
\begin{equation}
\lambda_\infty(\mu) \,\,=\,\, \mu^{d-4} \,
\left[ \frac{2}{d-4}\,  + \, \gamma_E \, - \,\ln 4\pi \,-\,1\, \right] \, .
\end{equation}
%%
%%\begin{equation}
%%\lambda_\infty = \frac{2\mu^{d-4}}{d-4} + \gamma_E - \ln 4\pi -1\, .
%%\end{equation}
%%

Still,  the Goldstone propagator and the Goldstone
decay amplitudes will be renormalized in the on-shell scheme,
as it is done in $\chi$PT, in order to ease the low-energy matching
of \rcht\ and \chpt\ at $\cO(p^2)$.  Everything else will be renormalized
in this section in $\overline{MS}-1$.
For simplicity, we will denote this set of schemes as
$\wms$ from now on.
Afterwards, we will study alternative renormalization schemes for the
\rcht\ couplings and their relation with the $\wms$  parameters.

In this section, together with the general structure of the
amplitudes, we will provide in this Section
just the explicit results for the case when
the lagrangian contains the operators $\mL_G+\mL_R$
with at most one resonance field,
derived by Ecker {\it et
al.}~\cite{Ecker1}.
The contributions from operators $\mL_{RR'}$ with two resonance fields
are provided separately later in  Sec.~\ref{sec.extra-operators}.
For clarity, we provide the
individual contributions from each absorptive cut (e.g.
$\pi\pi$, $V\pi$...). The precise definitions
for the corresponding Feynman integrals  are given in
Appendix~\ref{app.Feynman}.

\subsection{Goldstone boson renormalizations}

\subsubsection{Goldstone  self-energy}

\begin{figure}[t!]
\begin{center}
\epsfxsize=11cm\epsfbox{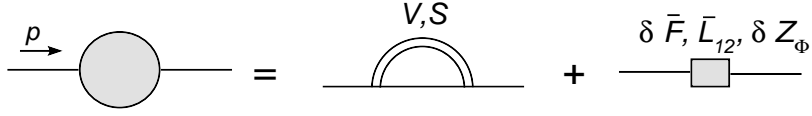}
\end{center}
\caption{\small Contributions to the Goldstone boson self-energy.
The single line represents the Goldstone boson while the double line
represents the resonance. The type of resonance is written above
it.} \label{fig.pi-self}
\end{figure}

The general form of the renormalized Goldstone propagator is given
by
\begin{equation}
i\Delta_\phi^{ -1} \,\, =\,\,  \frac{\widetilde{F}^2\, Z_\phi}{F^2}p^2
\,\,  -\,\, \frac{4 \widetilde{L}_{12}p^4}{F^2} \,\,
-\,\,  \Sigma_{\phi}(p^2) \, ,
\end{equation}
with $Z_\phi$ the wave function renormalization of the bare Goldstone field,
$\phi^{(B)}=Z_\phi^\frac{1}{2}\phi^r$.
In order to make the propagator finite, one needs to perform the  shifts
\begin{equation}
Z_\phi= 1 +\delta Z_\phi,\qquad \widetilde{F}=F + \delta
\widetilde{F},\qquad \widetilde{L}_{12} = \widetilde{L}_{12}^r  + \delta
\widetilde{L}_{12}\, ,
\end{equation}
where $\delta Z_\phi$ and $\delta \widetilde{F}$ are NLO in $1/N_C$.
The NLO coupling  $\widetilde{L}_{12}$ is split into a finite
renormalized part $\widetilde{L}^r_{12} $ and an infinite counter-term $\delta
\widetilde{L}_{12} $.

Considering the on-shell renormalization scheme for the Goldstone propagator,
i.e. such that $i \Delta^{-1}_\phi = p^2+{\cal O}(p^4)$,
leads to the renormalization condition
\begin{equation}
\frac{2\delta\widetilde{F}}{F} \, +\,  \delta Z_\phi
\, -\, \Sigma_\phi^{\,'}(0)
\,\,=\,\,0\, ,
\end{equation}
with $\Sigma_\phi^{\,'}(0) =\left.\Frac{d\Sigma_\phi }{dp^2}\right|_{p^2=0}$.
The ${\cal O}(p^4)$ ultraviolet divergence in $\Sigma_\phi $
is  absorbed into $\delta \widetilde{L}_{12}$ in the $\wms$ scheme.
The renormalized Goldstone propagator is then provided by
\begin{equation}
i\Delta_\phi^{ -1} =   p^2 \, -\, \Frac{ 4 \widetilde{L}^r_{12}  p^4}{F^2}
\, -\, \Sigma_\phi^r(p^2)  \, ,
\end{equation}
with its perturbative  expansion,
\begin{eqnarray}
\Delta_\phi^r &=& \frac{i}{p^2}
\, +\, \frac{i}{p^4}  \left[\Frac{4 \widetilde{L}^r_{12}  p^4}{F^2}
\, +\, \Sigma_\phi^r(p^2) \right]\,\,\,+\,\,\, ...
\end{eqnarray}
where \ the \  dots  \ stand  \  \ for the  next-to-next-to-leading
\ order \ corrections \  (NNLO) \  \ and \  \  ${ \Sigma_\phi^r(p^2)
=\Sigma_\phi(p^2)- p^2 \Sigma'_\phi(0)
-\Sigma_\phi(p^2)|_{\lambda_\infty \cO(p^4)}      }$ behaving like
$\cO(p^4)$ when $p^2\to 0$.

If one considers just the contributions $\mL_R$ from  interactions
linear in the resonance fields~\cite{Ecker1},
the one loop Goldstone self-energy $\Sigma_\phi$ is given by the
diagrams shown in Fig.~\ref{fig.pi-self}. A priori,
tadpole diagrams might appear, either with a Goldstone
or  a resonance running within
the  loop.
However,  they happen to be zero in the chiral limit.
All this yields the renormalizations and the renormalized self-energy,
\begin{eqnarray}
&&\frac{2\delta\widetilde{F}}{F} \, +\,  \delta Z_\phi
\,+\, \frac{1}{8F^4\pi^2}\left[\frac{9G_V^2M_V^2}{2}\left(\lambda_\infty+\ln\frac{M_V^2}{\mu^2}+\frac16\right)
-3c_d^2M_S^2\left(\lambda_\infty+\ln\frac{M_S^2}{\mu^2}-\frac12\right)\right]
\,\,=\,\,0\, ,
\nn\\
&&\delta \widetilde{L}_{12} \,\,=\,\,
-\frac{3(2c_d^2+G_V^2)}{64\pi^2F^2}\lambda_\infty\, ,
\end{eqnarray}
\begin{eqnarray}
\Sigma_\phi^r(p^2) |_{S\phi} &=&
 \frac{3 c_d^2 p^4}{8\pi^2 F^4}\left[\ln\frac{M_S^2}{\mu^2} +
\phi\left(\frac{p^2}{M_S^2}\right)\right]  \, ,
\nn\\
\Sigma_\phi^r(p^2) |_{V\phi} &=&
  \frac{3G_V^2 p^4}{16\pi^2F^4}\left[\ln\frac{M_V^2}{\mu^2}
+\phi\left(\frac{p^2}{M_V^2}\right)\right] \, ,
\label{eq.pion-self1}
\end{eqnarray}
with
\begin{eqnarray}
\phi(x) &=& \left(1-\frac{1}{x}\right)^3\ln(1-x)
-\frac{(x-2)\left(x-\frac12\right)}{x^2}\, .
\end{eqnarray}

\subsubsection{Vertex $p\phi$}

\begin{figure}[t!]
\begin{center}
\epsfxsize=11cm\epsfbox{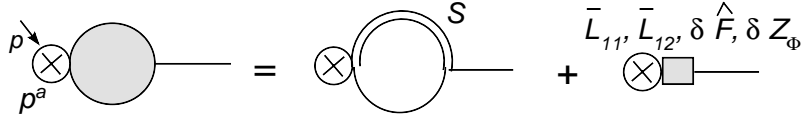}
\end{center}
\caption{\small Contribution to  the vertex $p\phi$. The crossed
circle stands for a pseudo-scalar density insertion.}
\label{fig.p-phi}
\end{figure}

The vertex function has the form
\begin{equation}
\Phi_{p\phi}(p^2)\,\, =\,\, \sqrt{2} \frac{Z_\phi^\frac{1}{2}\, \hat{F}^2\, B_0}{F}
-\frac{4\sqrt{2} B_0p^2}{F}(\widetilde{L}_{11}+\widetilde{L}_{12}) +
 \Phi_{p\phi}(p^2)^{1\ell}
\end{equation}
where $ \Phi_{p\phi}(p^2)^{1\ell}$ represents the one-particle-irreducible (1PI)
contribution  from meson loops.

Notice that it is convenient to choose the renormalization scheme
for $\delta \hat{F}$ such that the on-shell decay amplitude
coincides with the pion decay constant, which by construction we
denote as $F$. Thus, for the renormalizations
\begin{equation}
\hat{F}= F + \delta \hat{F},\qquad \widetilde{L}_{11} =
\widetilde{L}_{11}^r(\mu) + \delta \widetilde{L}_{11}(\mu)\, ,
\end{equation}
one has
\begin{eqnarray}
\Frac{ 2\delta \hat{F}}{F} \, +\,  \Frac{1}{2} \delta Z_\phi\,
  +\,  \Frac{1}{\sqrt{2} B_0 F} \Phi_{p\phi}(0)^{1\ell} \,\, =\,\, 0\, ,
\end{eqnarray}
and the counter-term $\delta \widetilde{L}_{11}(\mu)$  is chosen to
cancel the $\cO(p^2)$ divergent terms in $
\Phi_{p\phi}(p^2)^{1\ell}$ in the $\wms$--scheme. The renormalized
vertex function is then equal to
\begin{equation}
 \Phi_{p\phi}(p^2) \,\,=\,\, \sqrt{2} B_0 F
 \left\{\, 1
 \,- \,   \frac{4  \widetilde{L}_{11}^r p^2}{F^2}
 \,-\,    \frac{4 \widetilde{L}_{12}^r p^2}{F^2}
 \,+\,    \Frac{1}{\sqrt{2} B_0 F}\, \Phi_{p\phi}^r(p^2)^{1\ell}\,\right\} \, ,
\end{equation}
with $\Phi_{p\phi}^r(p^2)^{1\ell}$ being $\cO(p^2)$ when $p^2\to 0$.

In the case with  only $\mL_R$ interactions, linear in the resonance
fields~\cite{Ecker1},
one has the diagrams shown in Fig.~\ref{fig.p-phi}. These lead
to the renormalizations and renormalized one-loop contributions,
\begin{eqnarray}
\Frac{2\delta \hat{F}}{F} \, +\,  \Frac{1}{2} \delta Z_\phi &=& 0,\\
\delta \widetilde{L}_{11}(\mu) + \delta \widetilde{L}_{12}(\mu) &=&
-\frac{3c_dc_m}{16\pi^2F^2}\lambda_\infty\, ,
\end{eqnarray}
\begin{eqnarray}
 \Frac{1}{\sqrt{2} B_0 F} \Phi^r_{p\phi}(p^2)^{1\ell}|_{S\phi}  &=&
 \frac{3c_dc_m p^2}{4\pi^2F^4}\left[1-\ln\frac{M_S^2}{\mu^2} +
\psi\left(\frac{p^2}{M_S^2}\right)\right] \, ,
 \end{eqnarray}
with
\begin{eqnarray}
 \psi(x) &=& -\frac{1}{x} - \left(1-\frac{1}{x}\right)^2\ln(1-x)\, .
\end{eqnarray}

\subsubsection{Vertex $a\phi$}

\begin{figure}[t!]
\begin{center}
\epsfxsize=10cm\epsfbox{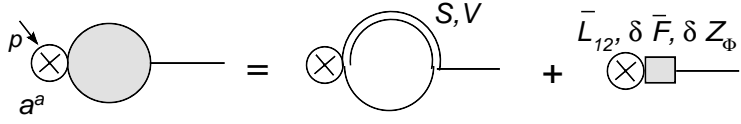}
\end{center}
\caption{\small Contribution to the vertex $p\phi$. The crossed
circle stands for a axial-vector current insertion.}
\label{fig.a-phi}
\end{figure}

Although it is not required for the correlator calculation in
this article, we will compute the $a\phi$ vertex function
for sake of completeness. From
previous calculations we obtained two equations for three
unknown objects $\delta \widetilde{F}$, $\delta \hat{F}$ and $\delta
Z_{\phi}$. The third equation can be found by analyzing the $a^\mu\to \phi$ vertex,
which, abusing of the notation,  has the form
\begin{equation}
 \Phi_{a\phi}(p)^\mu\,\, =\,\, \Phi_{a\phi}\,\cdot\,  p^\mu
\end{equation}
where
\begin{equation}
 \Phi_{a\phi}  = \sqrt{2} \frac{ \widetilde{F}^2\, Z_\phi^\frac{1}{2}}{F}
\,-\, \frac{4 \sqrt{2} \widetilde{L}_{12}p^2}{F} \,+\, \Phi_{a\phi}(p^2)^{1\ell}\, .
\end{equation}

As it happened before with $\delta \hat{F}$, it is convenient
to choose for $\delta \widetilde{F}$ (as we did here)
the scheme that recovers the  pion decay constant $F$
when the decay amplitude is set on-shell ($p^2\to 0$):
\begin{eqnarray}
\frac{2\delta \widetilde{F}}{F}\, +\, \Frac{1}{2} \delta Z_\phi
\,+\, \Frac{1}{\sqrt{2} F}\Phi_{a\phi}(0)^{1\ell}
\,\,=\,\, 0\, .
\end{eqnarray}
The coupling  $\delta \widetilde{L}_{12}(\mu)$ is chosen to cancel
the $\cO(p^2)$ UV divergent term  in $\Phi_{a\phi}(p^2)^{1\ell}$
in the $\wms$ scheme.

When  only $\mL_R$ interactions are taken into account~\cite{Ecker1},
the diagrams shown in Fig.~\ref{fig.a-phi} yield the renormalizations
\begin{equation}
\frac{2\delta \widetilde{F}}{F} + \Frac{1}{2}\delta Z_\phi
+\frac{1}{8\pi^2 F^4}\left[\frac{9G_V^2M_V^2}{2}\left(\lambda_\infty+\ln\frac{M_V^2}{\mu^2}+\frac16\right)
-3c_d^2M_S^2\left(\lambda_\infty+\ln\frac{M_S^2}{\mu^2}-\frac12\right)\right]\Bigg\}=
0
\end{equation}
\begin{equation}
\delta \widetilde{L}_{12}(\mu) \,=\, -
\frac{3(2c_d^2+G_V^2)}{64\pi^2F^2}\lambda_\infty\, .
\end{equation}
In this case, it is possible to see explicitly that the
renormalization for $\widetilde{L}_{12}$ is in an agreement with its
former result from the Goldstone propagator.

\subsubsection{Renormalization of $\hat{F}$, $\widetilde{F}$ and $\delta
Z_\phi$}

Comparing the three equations for $\delta\hat{F}$, $\delta\widetilde{F}$
and $\delta Z_\phi$, one is finally able to extract
each of them separately:
\begin{eqnarray}
\delta Z_\phi &=& 2 \Sigma_\phi^{\,'}(0)^{1\ell} \,
+ \, \Frac{\sqrt{2} }{F}\Phi_{a\phi}(0)^{1\ell}\, ,
\nn\\
\Frac{\delta \hat{F}}{F}  &=& -\Sigma_\phi^{\, '}(0)^{1\ell}\,
-\, \Frac{1}{\sqrt{2} B_0 F}\Phi_{p\phi}(0)^{1\ell}
\,-\, \Frac{1}{\sqrt{2} F}\Phi_{a\phi}(0)^{1\ell}\, ,
\nn\\
\Frac{\delta \widetilde{F}}{F}&=&   - \Frac{1}{2}\Sigma_\phi^{\,'}(0)^{1\ell}
\, -\, \Frac{1}{\sqrt{2} F} \Phi_{a\phi}(0)^{1\ell}\, .
\end{eqnarray}
Thus, in the case when only interactions $\mL_R$,
linear in the resonance fields,
are considered~\cite{Ecker1},
one gets  $\delta Z_\phi = 0$, $\delta\hat{F} = 0$ and
\begin{equation}
\delta\widetilde{F} =
-\frac{1}{16\pi^2F^4}\left[\frac{9G_V^2M_V^2}{2}\left(\lambda_\infty+\ln\frac{M_V^2}{\mu^2}+\frac16\right)
-3c_d^2M_S^2\left(\lambda_\infty+\ln\frac{M_S^2}{\mu^2}-\frac12\right)\right]\Bigg\}\, .
\end{equation}
This confirms the results from Ref.~\cite{Rosell-genera}, where
$\widetilde{F}$ was renormalized but $\hat{F}$ was not. On the other hand,
the renormalizations of $\hat{F}$ and $\widetilde{F}$ were not considered
in Ref.~\cite{L9} and, consequently, a nonzero $\delta Z_\phi$ was found.

\subsection{Scalar resonance renormalization}

\subsubsection{Scalar resonance self-energy}

\begin{figure}[t!]
\begin{center}
\epsfxsize=10cm\epsfbox{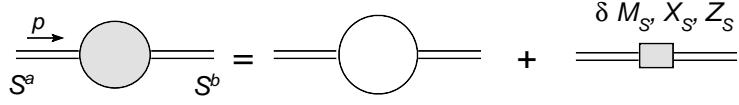}
\end{center}
\caption{\small Contributions to  the scalar resonance
self-energy}
\end{figure}

The renormalized propagator has the form
\begin{equation}
 i\Delta_S^{ -1} \,\,=\,\, Z_S\, (p^2\,-\, M_S^2)\, +\, X_S p^4 \,
-\, \Sigma_{S}(p^2) \, ,
\end{equation}
where we have performed the scalar resonance wave-function
renormalization $S^{(B)}=Z_S^\frac{1}{2} S^r$.
In order to cancel the $\lambda_\infty$ divergent terms of the
one-loop self-energy $\Sigma_S(p^2)$,
we make the shifts
\begin{equation}
M_S^2= M_S^{r\,\,2} +\delta M_S^2,\qquad Z_S = 1 + \delta Z_S,\qquad X_S =
X_S^r(\mu) + \delta X_S(\mu).
\end{equation}
The renormalized propagator is then given by,
\begin{equation}
i\Delta_S^{ -1} \,\,=\,\, p^2\,-\,M_S^{r\,\,2}\, +\,  X_S^r(\mu)p^4
\,-\, \Sigma_S^r(p^2) \,,
\end{equation}
with its perturbative expansion,
\begin{equation}
\Delta_S \,\, =\,\, \frac{i}{p^2-M_S^{r\,\, 2} }
\,\,+ \,\, \frac{i}{(p^2-M_S^{r\,\,2}   )^2}\Bigg\{
\,-\, X_S^r(\mu) p^4\, +\, \Sigma_S^r(p^2) \Bigg\}
+ \dots
\end{equation}

In the case where only the $\mL_R$ interactions are considered,
one obtains
\begin{equation}
\delta M_S = 0,\qquad\qquad \delta Z_S = 0,\qquad\qquad \delta
X_S(\mu) = \frac{3c_d^2}{16\pi^2F^4}\lambda_\infty\, .
\nn
\end{equation}
\begin{equation}
\Sigma_S^{r}(p^2) |_{\phi\phi}  \,\,=\,\,
-\, \frac{3c_d^2p^4}{16\pi^2F^4}\left[1-\ln\left(\frac{-p^2}{\mu^2}\right)\right]\,.
\label{eq.S-self1}
\end{equation}

\subsubsection{Vertex $sS$}

\begin{figure}[t!]
\begin{center}
\epsfxsize=10cm\epsfbox{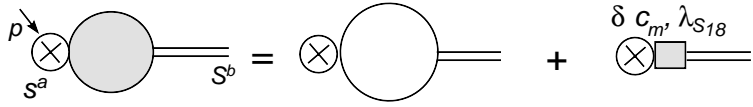}
\end{center}
\caption{\small Contributions to the  vertex $sS$. The crossed
circle stands for a scalar density insertion.}
\end{figure}

The vertex function $s(x)\to S$ has the form
\begin{equation}
\Phi_{sS}(p^2) \,\,=\,\,
-\,4 B_0\,\left\{\, Z_S^\frac{1}{2}\, c_m \,-\,  \lambda^S_{18}p^2
\,-\,\Frac{1}{ 4 B_0} \Phi_{sS}(p^2)^{1\ell}\, \right\}\,.
\end{equation}
The renormalizations of the scalar wave-function $Z_S=1+\delta Z_S$, the LO constant
$c_m=c_m^r(\mu)+\delta c_m(\mu)$  and  the NLO
coupling $\lambda^S_{18} = \lambda^S_{18}(\mu) + \delta
\lambda^S_{18}(\mu)$  make the amplitude finite:
\begin{equation}
\Phi_{sS}(p^2) \,\,=\,\,
-\,4 B_0\left\{ \, c_m^r \,-\,  \lambda^S_{18}(\mu) p^2
\,-\, \Frac{1}{  4 B_0} \Phi_{sS}^r(p^2)^{1\ell}\,\right\}\,.
\end{equation}

In the case with only $\mL_R$ interactions~\cite{Ecker1},
we had $\delta Z_S=0$.
The cancelation of divergences in the $\wms$ scheme leads to the
shift and the renormalized one-loop contributions,
\begin{equation}
\delta c_m=0\, ,\qquad\qquad \qquad \delta \lambda^S_{18}(\mu) =
-\frac{3c_d}{64\pi^2F^2}\lambda_\infty
\end{equation}
\begin{equation}
-\Frac{1}{ 4 B_0} \Phi^r_{sS}(p^2)^{1\ell} |_{\phi\phi}\,\, =\,\,
\frac{3c_dp^2}{64\pi^2F^2}\left(1-\ln\frac{-p^2}{\mu^2}\right) \, .
\label{eq.sS-vertex1}
\end{equation}

\subsection{Pseudo-scalar resonance renormalization}

\subsubsection{Pseudoscalar resonance self-energy}

\begin{figure}[t!]
\begin{center}
\epsfxsize=6cm\epsfbox{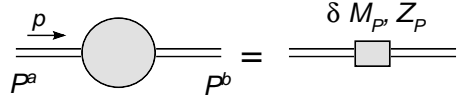}
\end{center}
\caption{\small Contribution to the   pseudoscalar
resonance self-energy}
\end{figure}

The renormalized  pseudoscalar propagator has the form
\begin{equation}
 i\Delta_P^{ -1} \,\,=\,\, Z_P\, (p^2\,-\, M_P^2)\, +\, X_P p^4 \,
-\, \Sigma_{P}(p^2) \, ,
\end{equation}
with  $P^{(B)}=Z_P^\frac{1}{2} P^r$.
The cancelation of the $\lambda_\infty$ UV divergent terms in
the one-loop self-energy $\Sigma_P(p^2)$ needs the shifts
\begin{equation}
M_P^2= M_P^{r\,\,2} +\delta M_P^2,\qquad Z_P = 1 + \delta Z_P,\qquad X_P =
X_P^r(\mu) + \delta X_P(\mu),
\end{equation}
leading to the renormalized propagator,
\begin{equation}
i\Delta_P^{ -1} \,\,=\,\, p^2\,-\,M_P^{r\,\,2}\, +\,  X_P^r(\mu)p^4
\,-\, \Sigma_P^r(p^2) \,,
\end{equation}
and its perturbative expansion,
\begin{equation}
\Delta_P \,\, =\,\, \frac{i}{p^2-M_P^{r\,\, 2} }
\,\,+ \,\, \frac{i}{(p^2-M_P^{r\,\,2}   )^2}\Bigg\{
\,-\, X_P^r(\mu) p^4\, +\, \Sigma_P^r(p^2)   \Bigg\}
+ \dots
\end{equation}

In the case where only the $\mL_R$ interactions are considered~\cite{Ecker1},
there is no one-loop diagrams contributing and,
therefore, $\delta Z_P=\delta M_P^2=\delta X_P=0$.

\subsubsection{Vertex $pP$}

\begin{figure}[t!]
\begin{center}
\epsfxsize=5.5cm\epsfbox{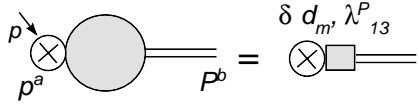}
\end{center}
\caption{Contribution to the renormalization of vertex $pP$}
\end{figure}

The vertex function $p(x)\to P$ has the form
\begin{equation}
\Phi_{pP}(p^2) \,\,=\,\,
-\, 4  B_0\,\left\{\, Z_P^\frac{1}{2}\, d_m \,-\,  \lambda^P_{13}p^2
\,-\,\Frac{1}{  4 B_0} \Phi_{pP}(p^2)^{1\ell}\, \right\}\,.
\end{equation}
The renormalizations of the scalar wave-function $Z_P=1+\delta Z_P$, the LO constant
$d_m=d_m^r(\mu)+\delta d_m(\mu)$  and  the NLO
coupling $\lambda^P_{13} = \lambda^P_{13}(\mu) + \delta
\lambda^P_{13}(\mu)$  make the amplitude finite:
\begin{equation}
\Phi_{pP}(p^2) \,\,=\,\,
-\, 4  B_0\left\{ \, d_m^r \,-\,  \lambda^P_{13}(\mu) p^2
\,-\, \Frac{1}{  4  B_0} \Phi_{pP}^r(p^2)^{1\ell}\,\right\}\,.
\end{equation}

In the case with only $\mL_R$ interactions~\cite{Ecker1},  $\delta Z_P=0$ and
there is no loop diagram contributing to this vertex, so we have
$\delta d_m=\delta \lambda_{13}^P=0$ and the renormalized vertex function
results
\begin{equation}
\Phi_{pP}(p^2) \,\,=\,\, -4 B_0 \, \left\{ d_m^r \, - \,
\lambda^P_{13}(\mu) \,    p^2\, \right\}\, .
\end{equation}

\subsection{1PI contributions}

\subsubsection{1PI diagram $ss$}

\begin{figure}[t!]
\begin{center}
\epsfxsize=11cm\epsfbox{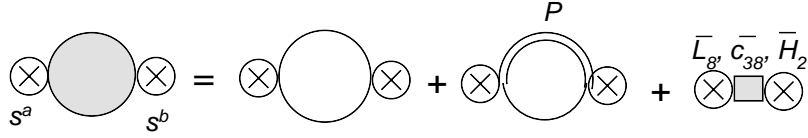}
\end{center}
\caption{\small Contribution to the   1PI vertex $ss$}
\label{fig.ss}
\end{figure}

Now, we analyze 1PI diagrams that appear in the $ss$--correlator:
\begin{equation}
\Pi^{\rm 1PI}_{ss}(p^2) \,\, =\,\,
16 B_0^2 \widetilde{L}_8
\,+\,8 B_0^2 \widetilde{H}_2
%%
%%\,+\,32 B_0^2 \widetilde{C}_{38}p^2
%%
\,+\,\Pi_{ss}^{\rm 1PI}(p^2)^{1\ell}\, .
\end{equation}
The shifts
$\widetilde{L}_8 = \widetilde{L}_8^r(\mu) + \delta \widetilde{L}_8(\mu)$
and
$\widetilde{H}_2 = \widetilde{H}_2^r(\mu) + \delta \widetilde{H}_2(\mu)$
%%
%%$\widetilde{C}_{38}=\widetilde{C}_{38}^r(\mu)+\delta \widetilde{C}_{38}(\mu)$
%%
render the amplitude finite by canceling the UV divergences in the $\wms$--scheme,
which becomes
\begin{equation}
\Pi^{\rm 1PI}_{ss}(p^2) \,\, =\,\,
16 B_0^2 \widetilde{L}^r_8(\mu)
\,+\,8 B_0^2 \widetilde{H}^r_2(\mu)
%%
%%\,+\,32 B_0^2 \widetilde{C}^r_{38}(\mu) p^2
%%
\,+\,\Pi_{ss}^{\rm 1PI, \, r}(p^2)^{1\ell}\, .
\end{equation}

In the case with only interactions $\mL_R$ linear in the resonance
fields~\cite{Ecker1}, the 1PI
 diagrams contributing to the   $SS$--correlator are
shown in Fig.~\ref{fig.ss}.  Thus, one gets for the shifts and the
renormalized amplitude the expressions
\begin{equation}
2\delta\widetilde{L}_8(\mu)+\delta \widetilde{H}_2(\mu) =
\frac{3(F^2+16d_m^2)}{128\pi^2F^2}\lambda_\infty\, ,
%%
%%\qquad\qquad\qquad \delta \widetilde{C}_{38}=0\, ,
%%
\end{equation}
\begin{eqnarray}
\Pi^{\rm 1PI,\, r}_{ss}(p^2)^{1\ell}|_{\phi\phi} \,\, &=&\,\,  B_0^2\,\,\Frac{3}{16\pi^2}\,
\left[\, 1\, -\, \ln{\Frac{-p^2}{\mu^2}} \, \right]\, ,
\nn\\
\Pi^{\rm 1PI,\, r}_{ss}(p^2)^{1\ell}|_{P \phi} \,\, &=&\,\,B_0^2\,\,
 \Frac{3 d_m^2}{\pi^2 F^2}\,
\left[  1\, -\,  \ln\frac{M_P^2}{\mu^2}\, -\, \left(1-\Frac{M_P^2}{p^2}\right)
\ln{\left(1-\Frac{p^2}{M_P^2}\right)  }    \right]\, .
\end{eqnarray}

\subsubsection{1PI diagram $pp$}

\begin{figure}[t!]
\begin{center}
\epsfxsize=8.5cm\epsfbox{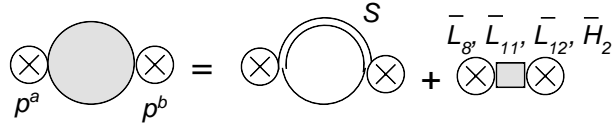}
\end{center}
\caption{\small Contribution to the  1PI vertex $pp$}
\end{figure}

Similarly, for $pp$--amplitude one has the structure,
\begin{equation}
\Pi^{\rm 1PI }_{pp}(p^2) \,\,=\,\,
-\, 16 B_0^2 \widetilde{L}_8
\,-\, 16 B_0^2 \widetilde{L}_{11}
\,-\, 8 B_0^2 \widetilde{L}_{12}
\, +\, 8 B_0^2 \widetilde{H}_2
\,\, +\,\, \Pi^{\rm 1PI }_{pp}(p^2) ^{1\ell}\, .
\end{equation}
The UV divergences are absorbed through the renormalization of
$\widetilde{L}_8$, $\widetilde{L}_{11}$, $\widetilde{L}_{12}$
and $\widetilde{H}_{12}$, rendering the amplitude finite:
\begin{equation}
\Pi^{\rm 1PI }_{pp}(p^2) \,\,=\,\,
-\, 16 B_0^2 \widetilde{L}_8^r(\mu)
\,-\, 16 B_0^2 \widetilde{L}_{11}^r(\mu)
\,-\, 8 B_0^2 \widetilde{L}_{12}^r(\mu)
\, +\, 8 B_0^2 \widetilde{H}_2^r(\mu)
\,\, +\,\, \Pi^{\rm 1PI, \, r}_{pp}(p^2)^{1\ell}\, .
\end{equation}

In the case where only the contributions from $\mL_R$ operators are
considered~\cite{Ecker1}, the divergences are absorbed by the  shift
\begin{equation}
2\delta \widetilde{L}_8(\mu) +2 \delta \widetilde{L}_{11}(\mu) +\delta
\widetilde{L}_{12}(\mu)-\delta\widetilde{H}_2 =
-\frac{3c_m^2}{8\pi^2F^2}\lambda_\infty\, ,
\end{equation}
leaving the finite one-loop contribution,
\begin{equation}
\Pi^{\rm 1PI, \, r}_{pp}(p^2)^{1\ell}|_{S\phi}
\,\, = \,\,  B_0^2  \,\,
\Frac{3c_m^2}{\pi^2 F^2}\, \left[1\, -\, \ln\frac{M_S^2}{\mu^2}
\, -\, \left(1-\Frac{p^2}{M_S^2}\right)\ln{\left(1+\Frac{p^2}{M_S^2}\right)}  \right]
\, .
\end{equation}

\subsection{Correlator at NLO}

At NLO we can write the general 1PI decomposition of the $SS-PP$ correlator
in terms of  renormalized  correlators and vertex functions,
\begin{eqnarray}
\Pi_{ss-pp}(p^2)&=&
\, i\,   \Delta_S (p^2)\,\,\left\{\, \Phi_{sS} (p^2)\,\right\}^2
\,\,\,\,  -\,\,\,\,  i \, \Delta_P (p^2)\, \left\{\,\Phi_{pP}(p^2)\,\right\}^2
\,\, \,\, -\,\,\,\,  i\, \Delta_\phi (p^2)\, \left\{\, \Phi_{p\phi}(p^2)\,\right\}^2
\nn\\
\nn\\
&&\quad +\,\,\Pi_{ss}^{\rm 1PI}(p^2)\,\,\,\, -\,\,\,\, \Pi_{pp}^{\rm 1PI}(p^2)\, ,
\label{eq.general-PI}
\end{eqnarray}
where we made use of the relation between the vertex functions for incoming and
outgoing mesons,
$\Phi_{sS} =\Phi_{Ss} $, $\Phi_{pP} =\Phi_{Pp} $,
$\Phi_{p\phi} =\Phi_{\phi p} $.

If one now uses the previous perturbative calculation,
the $SS-PP$ octet correlator takes up to NLO in $1/N_C$ the form,
\begin{eqnarray}
\Frac{1}{B_0^2}\, \Pi(p^2)&=&
\Frac{1}{M_S^2 -p^2}\left(16 c_m^2- 32 c_m \lambda_{18}^S p^2
+\Frac{16 c_m^2X_S p^4}{M_S^2-p^2}\right)
\nn\\
&&\qquad \qquad\qquad
-\Frac{ 16 c_m^2}{(M_S^2-p^2)^2}\Sigma_S^r(p^2)^{1\ell}
-\Frac{ 8 c_m}{M_S^2-p^2}\,\Frac{1}{B_0}\Phi_{sS}^r(p^2)^{1\ell}
\nn\\
\nn\\
&&
- \,\Frac{1}{M_P^2 -p^2}\left(16 d_m^2- 32 d_m \lambda_{13}^P p^2
+\Frac{16 d_m^2 X_P p^4}{M_P^2-p^2}\right)
\nn\\
&&\qquad \qquad\qquad
+\Frac{ 16 d_m^2}{(M_P^2-p^2)^2}\Sigma_P^r(p^2)^{1\ell}
+\Frac{ 8 d_m}{M_P^2-p^2}\,\Frac{1}{B_0}\Phi_{pP}^r(p^2)^{1\ell}
\nn\\
\nn\\
&&
+\, \Frac{2 F^2}{p^2}\left(1- \Frac{8\widetilde{L}_{11} p^2}{F^2}
-\Frac{4 \widetilde{L}_{12} p^2}{F^2} \right)
+\Frac{2 F^2}{p^4}\Sigma_\phi^r(p^2)^{1\ell}
+\Frac{2 F}{p^2}\,\Frac{\sqrt{2}}{B_0}\Phi_{p\phi}^r(p^2)^{1\ell}
\nn\\
&&
+ 32 \widetilde{L}_8+16\widetilde{L}_{11} +8\widetilde{L}_{12}
%%
%%+32 \widetilde{C}_{38}p^2
%%
\,\,\,\, +\,\,\,\,\Pi_{ss-pp}^r(p^2)^{1\ell}\, .
\label{eq.NLO-correl}
\end{eqnarray}

The couplings shown here (and from now on) are the renormalized ones
even if the superscript ``$r$''  is not explicitly present. The
first two lines are the contribution from the scalar exchanges. The
third and fourth ones come from the pseudoscalar resonance
exchanges, whereas the fifth one is produced by the Goldstone
exchanges. The last line is given by the 1PI diagrams in the $SS-PP$
correlator.

Notice that the correlator results independent of $\widetilde{L}_{11}$ and
$\widetilde{L}_{12}$ due to the cancelation between
the Goldstone exchanges and the 1PI terms in~(\ref{eq.NLO-correl}).
Likewise, it is possible to  check that the correlator
only depends on the effective
combinations $c_m^{\rm eff}$, $d_m^{\rm eff}$, $M_S^{\rm eff}$, $M_P^{\rm eff}$
$\widetilde{L}_8^{\rm eff}$
%%
%%and $\widetilde{C}_{38}^{\rm eff}$
%%
from Eq.~(\ref{eq.eff-couplings}):
\begin{eqnarray}
\Frac{1}{B_0^2}\, \Pi(p^2)&=&
\Frac{16 c_m^{{\rm eff}\,\,2}}{M_S^{{\rm eff}\,\, 2} -p^2}
-\Frac{ 16 c_m^2}{(M_S^2-p^2)^2}\Sigma_S^r(p^2)^{1\ell}
-\Frac{ 8 c_m}{M_S^2-p^2}\,\Frac{1}{B_0}\Phi_{sS}^r(p^2)^{1\ell}
\nn\\
\nn\\
&&
- \,\Frac{16 d_m^{{\rm eff}\,\, 2}}{M_P^{{\rm eff}\,\, 2} -p^2}
+\Frac{ 16 d_m^2}{(M_P^2-p^2)^2}\Sigma_P^r(p^2)^{1\ell}
+\Frac{ 8 d_m}{M_P^2-p^2}\,\Frac{1}{B_0}\Phi_{pP}^r(p^2)^{1\ell}
\nn\\
\nn\\
&&
+\, \Frac{2 F^2}{p^2}
+\Frac{2 F^2}{p^4}\Sigma_\phi^r(p^2)^{1\ell}
+\Frac{2 F}{p^2}\,\Frac{\sqrt{2}}{B_0}\Phi_{p\phi}^r(p^2)^{1\ell}
\nn\\
&&
+ 32 \widetilde{L}_8^{{\rm eff}\,\, 2}
%%
%%+32 \widetilde{C}_{38}^{{\rm eff}\,\, 2} p^2
%%
\,\,\,\, +\,\,\,\,\Pi_{ss-pp}^r(p^2)^{1\ell}\, .
\label{eq.NLO-correl2}
\end{eqnarray}
The couplings   $X_S$, $X_P$, $\lambda_{18}^S$ and $\lambda_{13}^P$
disappear from our NLO calculation and $c_m $, $d_m $, $M_R $ and
$\widetilde{L}_8 $ are  replaced everywhere by
$c_m^{\rm eff}$, $d_m^{\rm eff}$, $M_R^{\rm eff}$ and
$\widetilde{L}_8^{\rm eff}$.  The replacement in the subleading
terms  leaves the expression unaltered up to the order in $1/N_C$
considered in our  computation.

This elimination  of the renormalized couplings
$X_S$, $X_P$, $\lambda_{18}^S$ and
$\lambda_{13}^P$ can be understood in an equivalent way by means of the EOM
of the theory and the meson field redefinitions.
The effective couplings that are left in front of the operators
after the meson field transformations coincide exactly with
the combinations that determine the correlator up to NLO.

In the subleading terms in Eq.~(\ref{eq.NLO-correl2}), a priori
one can use indistinct the original couplings, e.g. $c_m$,
or the effective ones, this is, $c_m^{\rm eff}$, as the difference
goes to NNLO.
However, for sake of consistence, one should always consider
the same renormalized coupling everywhere  in the amplitude.
Hence, after performing the field redefinition that removes
$X_S$, $X_P$, $\lambda_{18}^S$ and $\lambda_{13}^P$,
all the remaining couplings appearing in $\Pi(p^2)$
are the effective ones. From now on,
we will consider  that the \rcht\ action has been simplified through
meson field redefinitions in the previous way
and the superscript ``eff''   will be implicitly assumed in the couplings
in order to make the notation simpler.

\section{High energy constraints}
\label{sec.high}

The NLO expression for the correlator contains plenty of resonance
parameters that are not fully well known. A typical  procedure to
improve the determination of these couplings is the use of the
short-distance conditions~\cite{Ecker2}.

The operator product expansion tells us that the $SS-PP$ correlator
vanishes like $1/p^4$ for the large Euclidean momentum. Indeed, due
to the smallness of its dimension--four condensate
($\frac{1}{B_0^2}\bra \mO_4^{SS-PP}\ket \simeq  12\pi\alpha_S
F^4\sim 3 \cdot 10^{-4}$~GeV$^4$~\cite{Peris-O4}), it is a good
approximation to consider that it vanishes like $1/p^6$ when $p^2\to
-\infty$~\cite{PI:08,Peris-O4}.

The \rcht\  correlator does not follow this short-distance behaviour for arbitrary values of
its couplings. This imposes severe constraints on the coefficients
of the high-energy expansion of our NLO correlator,
\begin{equation}
\Frac{1}{B_0^2} \, \Pi(p^2) \,\,\,=\,\,\,
\sum_{n=0,1,2...}  \Frac{1}{(p^2)^k}\,
\left(\alpha_{2n}^{(p)}\, +\, \alpha_{2n}^{(\ell)}\ln \frac{-p^2}{\mu^2}
 \right)\, .
\end{equation}
The proper OPE short-distance behaviour is therefore recovered
by demanding~\cite{NLO-satura}
\begin{equation}
\alpha_{k}^{(\ell)}=\alpha_k^{(p)}=0\, , \qquad\mbox{for}\quad
k=0,2,4\, .
\end{equation}

At large $N_C$, there are no logarithmic terms
($\alpha_k^{(\ell)}=0$) and for the remaining coefficients one has
$\alpha_0^{(p)}=0$ (no $\widetilde{L}_8$ or higher local  couplings
at large $N_C$) and the two Weinberg sum-rules (WSR)~\cite{PI:08},
\begin{eqnarray}
\alpha_2^{(p)}  &=& 2F^2+ 16 d_m^2-16c_m^2   \,\,\,\, =\,\,\,\, 0 \,,
\nonumber\\
\alpha_4^{(p)} &=&
16 d_m^2 M_P^{2}  -16 c_m^2 M_S^2    \,\,\,\, =\,\,\,\, 0 \, .
\label{eq.largeNc-WSR}
\end{eqnarray}

At NLO,  in the case when the interactions only contain operators $\mL_R$
with at most one resonance field~\cite{Ecker1},
the high-energy expansion log--term coefficients result
\begin{eqnarray}
\Frac{8\pi^2 F^2}{3}\alpha_0^{(\ell)} &=& 8 c_m^2 - 8d_m^2 -4c_dc_m +2c_d^2 + G_V^2 -\frac{8c_d^2c_m^2}{F^2}
-\frac{F^2}{2}  \, ,
\nn\\
\Frac{8\pi^2 F^2}{3}\alpha_2^{(\ell)} &=& 8d_m^2M_P^2-8c_m^2M_S^2 -\frac{16M_S^2c_d^2c_m^2}{F^2}
+20c_dc_mM_S^2 -6c_d^2M_S^2 -3G_V^2M_V^2 \,,
\nn\\
\Frac{8\pi^2 F^2}{3}\alpha_4^{(\ell)} &=&
-\frac{24c_d^2c_m^2M_S^4}{F^2}-4c_dc_mM_S^4+6c_d^2M_S^4+3G_V^2M_V^4\,,
\label{eq.Ecker-log-const}
\end{eqnarray}
and the  high-energy coefficients $\alpha_{0,2,4}^{(p)}$ are given by
\begin{eqnarray}
\alpha_0^{(p)}   &=&  -\alpha_0^{(l)}\,\,\,\, +\,\,\,\, 32\, \widetilde{L}_8\, ,
\nn\\
\alpha_2^{(p)}  &=& 2F^2+ 16 d_m^2-16c_m^2 \,\,\,
+\,\,\, A(\mu)   \,,
\nn\\
\alpha_4^{(p)} &=&
16 d_m^2 M_P^{2}  -16 c_m^2 M_S^2
\,\,\,+\,\,\, B(\mu) \, ,
\end{eqnarray}
with the NLO corrections
\begin{eqnarray}
A(\mu)  &=&- \frac{3d_m^2M_P^2}{\pi^2F^2}\left(\ln\frac{M_P^2}{\mu^2}-1\right)
+\frac{3c_m^2M_S^2}{\pi^2F^2}\left(\ln\frac{M_S^2}{\mu^2}-1\right)
+\frac{6c_d^2c_m^2M_S^2}{\pi^2F^4}
\nn\\
&& \qquad
-\frac{6c_dc_mM_S^2}{\pi^2F^2}\left(\ln\frac{M_S^2}{\mu^2}+\frac14\right)
+\frac{9c_d^2M_S^2}{4\pi^2F^2}\left(\ln\frac{M_S^2}{\mu^2}+\frac12\right)
+\frac{9G_V^2M_V^2}{8\pi^2F^2}\left(\ln\frac{M_V^2}{\mu^2}+\frac12\right)  \,,
\nn\\
B(\mu)&=&
-\frac{3d_m^2M_P^4}{2F^2\pi^2} +\frac{9c_d^2c_m^2M_S^4}{F^4\pi^2}
+\frac{3c_m^2M_S^4}{2F^2\pi^2}-\frac{6c_dc_mM_S^4}{\pi^2F^2}
-\frac{9c_d^2M_S^4}{4\pi^2F^2}\left(\ln\frac{M_S^2}{\mu^2}-\frac12\right)
\nonumber\\
&&\qquad
+\frac{3c_dc_mM_S^4}{\pi^2F^2}\ln\frac{M_S^2}{\mu^2}
-\frac{9G_V^2M_V^4}{8\pi^2F^2}\left(\ln\frac{M_V^2}{\mu^2}- \frac12\right)
\, .  \label{eq.A+B}
\end{eqnarray}
The large--$N_C$ WSR~(\ref{eq.WSR}) gain the subleading
contributions in $1/N_C$, yielding for
$\alpha_2^{(p)}=\alpha_4^{(p)}=0$ the solution~\footnote{ The
notation $A(\mu)= 2 F^2 \delta^{(1)}_{NLO}$, $2 F^2 M_S^2
\delta_{NLO}^{(2)}=B(\mu)$ was used in Ref.~\cite{L8} }~\cite{L8},
\begin{eqnarray}
c_m^2\, =\, \Frac{F^2}{8}\, \Frac{M_P^2}{M_P^2-M_S^2}\left(1\, + \, \frac{A(\mu)}{2F^2}
-\frac{B(\mu)}{2 F^2  M_P^2}\right) ,
\,\,
d_m^2\, =\, \Frac{F^2}{8}\, \Frac{M_S^2}{M_P^2-M_S^2}\left(1\, + \, \frac{A(\mu)}{2 F^2}
-\frac{B(\mu)}{2F^2 M_S^2}\right)  .
\nn\\
\label{eq.NLO-WSR}
\end{eqnarray}
The couplings  $M_S$, $M_P$, $c_m$ and $d_m$ may also depend on $\mu$.
Nonetheless, unless necessary, this dependence will not be explicitly shown.
Also, as stated at the end of the previous section, one must keep in mind
that these are the results after the meson field redefinition that removes
the redundant couplings  $X_{S},X_{P},\lambda_{18}^S, \lambda_{13}^P$,
so the surviving couplings carry the superscript ``eff'' implicit.

The $\alpha_0^{(\ell)}=0$ constraint implies that
$\widetilde{L}_8=0$ also at NLO in $1/N_C$   (for any
renormalization scale $\mu$). We will see that for all the possible
interactions considered in this  paper, now here and later on, there
is the same constraint $\alpha_0^{(p)} = -\alpha_0^{(\ell)}  + 32
\widetilde{L}_8$ and, therefore, in general we find
$\widetilde{L}_8=0$.  The constants
$\alpha_k^{(\ell)}$, $A(\mu)$ and $B(\mu)$ only arise at NLO or higher.
Hence, when they are used for the computation of the correlator up
to NLO, one can indistinctly use for their calculation either
renormalized couplings or their large--$N_C$ values, as the
difference goes to NNLO.

Although these expressions  will be used later in other renormalization schemes,
$A(\mu)$ and $B(\mu)$ will always refer to their
former definitions in the $\wms$ scheme, like, for instance, the
results provided in Eq.~(\ref{eq.A+B}).

\subsection{Alternative renormalization schemes}

During the renormalization procedure we considered the
$\wms$--subtraction-scheme for all the resonance couplings. However,
in some situations one may get large contributions from $A(\mu)$ and
$B(\mu)$. The NLO prediction for $c_m$ and $d_m$ derived from
Eq.~(\ref{eq.NLO-WSR}) may then become very different from the
large--$N_C$ WSR determinations
$c_m^2=\frac{F^2}{8}\frac{M_P^2}{M_P^2-M_S^2}$,
$d_m^2=\frac{F^2}{8}\frac{M_S^2}{M_P^2-M_S^2}$.

A way out to minimize possible large radiative corrections to the
WSR is the choice of convenient renormalization schemes for
couplings ($c_m$ and $d_m$) and masses ($M_S$ and $M_P$). In the
renormalization procedure we originally chose to cancel the
$\lambda_\infty$   from the one-loop diagrams, but we could have
chosen to cancel the $\lambda_\infty$ term plus an arbitrary subleading
constant. This
change makes  that instead of having in the amplitudes the
renormalized coupling $\lambda^r_{\# 1}$ in the first scheme, one
now has the renormalized coupling in the second scheme plus a
constant, $\lambda^r_{\# 2}+C^{^{\# 1\to\# 2}}$. Thus, effectively
one can account for a change from the
$\wms$--subtraction-scheme (with renormalized couplings
$\kappa=c_m,d_m,M_S^2,M_P^2$) to another (with parameters
$\hat{\kappa}=\hat{c}_m,\hat{d}_m,\hat{M}_S^2,\hat{M}_P^2$) through
the shifts,
\begin{eqnarray}
\kappa\,\,\, =\,\,\, \hat{\kappa}\,\,+\,\, \Delta\kappa\, .
\end{eqnarray}
The difference $\Delta\kappa$ will  be, of course, subleading in the $1/N_C$
counting with respect to $\kappa$ and $\hat{\kappa}$.
This will affect the parts of the calculation where these couplings
contribute at LO in $1/N_C$. In the contributions that start at NLO
(e.g. $\alpha_k^{(\ell)}$, $A(\mu)$ and $B(\mu)$), the
variations due to $\Delta \kappa$ go to NNLO and they are
therefore neglected.
If one applies this change of scheme to Eq.~(\ref{eq.NLO-WSR}),
one gets for the NLO extension of the WSR,
\begin{eqnarray}
\alpha_2^{(p)}  &=& 2F^2+ 16 \, \hat{d}_m^2-16 \, \hat{c}_m^2 \,\,\,
+\,\,\,\left( 32  \, \hat{d}_m \Delta d_m -32  \, \hat{c}_m \Delta c_m\,\,+\,\,  A(\mu)
\right) \,,
\nn\\
\alpha_4^{(p)} &=&
16  \, \hat{d}_m^2 \hat{M}_P^{2}  -16 \,  \hat{c}_m^2 \hat{M}_S^2
\\
&&\qquad\,\,
\,\,\,+\,\,\, \left(  32  \, \hat{M}_P^2 \hat{d}_m \Delta d_m  + 16 \,  \hat{d}_m^2 \Delta M_P^2
-32  \, \hat{M}_S^2 \hat{c}_m \Delta c_m -  16 \,  \hat{c}_m^2 \Delta M_S^2\,\,+\,\,  B(\mu)
\right)  \, .
\nn
\end{eqnarray}
The terms within the brackets,  $( \cdots )$,
would be the finite contributions from the one-loop diagrams in the new scheme.

\subsubsection{Pole mass scheme for $M_S$ and $M_P$}

In addition to the $\wms$--scheme for the scalar and pseudo-scalar masses
($\Delta M_R^2=0$),
we will also study the pole--mass scheme.
The problem with the $\wms$ mass is the difficulty  to give a direct
physical meaning to the $\mu$--dependent mass  $M_R(\mu)$,
specially when  more and more  operators are added to the \rcht\ action.
On the other hand, the resonance pole mass is a universal property
which does not rely on any particular lagrangian realization.
Thus, instead  of considering the
$\mu$--dependent renormalized masses $M_R(\mu)$, we will switch to the renormalization
scale independent pole masses $\hat{M}_R=M_R^{^{\rm pole}}$,
defined by the pole positions
$(M_R^{^{\rm pole}}-i\Gamma_R^{^{\rm pole}}/2)^2$
of the renormalized propagators. Up to NLO in $1/N_C$,  one has
\begin{equation}
M_R^{^{\rm pole}\,2}\,\,=\,\, M_R^2\, +\, \mbox{Re} \Sigma_R^r(M_R^2)  \, ,
\qquad\qquad \qquad
M_R^{^{\rm pole}}\Gamma_R^{^{\rm pole}}\,\,=\,\,- \, \mbox{Im} \Sigma_R^r(M_R^2)  \, ,
\label{eq.pole-pos}
\end{equation}
and therefore,
\begin{equation}
\Delta M_R^2\,\,=\,\,  M_R^2 -\hM_R^2
\,\, =\,\,   -\mbox{Re} \Sigma_R^r(M_R^2)\, ,
\end{equation}
Since $\Delta M_R^2$ is NLO in $1/N_C$, the difference between
using the  $M_R^2$ ($\wms$--subtraction-scheme) within $\Sigma(M_R^2)$
or its value $\hM_R$ in another scheme goes to NNLO. Therefore, it is negligible at the perturbative order
we are working at.

If only interactions $\mL_R$ given by operators linear in the
resonance fields are taken into account~\cite{Ecker1}, one has for the pole scheme
\begin{eqnarray}
\Delta M_S^2\,\, =\,\,
\Frac{ 3 c_d^2 M_S^4}{16\pi^2 F^4}\left[1-\ln\Frac{M_R^2}{\mu^2}\right]\, ,
\qquad \qquad\qquad
\Delta M_P^2\,\,=\,\, 0\, ,
\end{eqnarray}
where only the two--Goldstone loop
$\Sigma_S(p^2)|_{\phi\phi}$ contributes to $\Delta M_S^2$  and
$\Sigma_P(p^2)=0$ if  only the $\mL_R$ interactions are taken into
account~\cite{Ecker1}.

\subsubsection{WSR--scheme for $c_m$ and $d_m$}

Since the value of the spin--0 parameters is very poorly known at
the experimental level,  one finds important uncertainties and variations
in the determination of $c_m$ and $d_m$ through
the NLO sum-rules~(\ref{eq.NLO-WSR}).
The choice of a shift that minimizes the finite part
of the loop contributions is not straight-forward.
For instance, within the $\wms$--subtraction-scheme itself,
it is not easy to find a value of $\mu$ that minimizes both $A(\mu)$ and
$B(\mu)$ at once unless the resonance couplings are appropriately fine-tuned.
This makes the short-distance matching rather cumbersome and the
extraction of the necessary resonance parameters problematic.

Alternatively, the selection of a shift $\Delta\kappa$ that exactly cancels
the one-loop contributions to Eq.~(\ref{eq.NLO-WSR})
(provided in the $\wms$--scheme by the constants $A(\mu)$ and $B(\mu)$)
seems to be a better option.
This converts Eq.~(\ref{eq.NLO-WSR}) into
\begin{equation}
\alpha_2^{(p)}\,=\, 2F^2+ 16 \,\hat{d}_m^2-16\,\hat{c}_m^2 \,=\, 0\,, \qquad \qquad
\alpha_4^{(p)} \,=\,
16 \,\hat{d}_m^2 \hat{M}_P^{2}  -16 \,\hat{c}_m^2 \hat{M}_S^2\,=\, 0\,,
\end{equation}
with the solutions
\begin{equation}
\hat{c}_m^2=\frac{F^2}{8}\frac{\hat{M}_P^2}{\hat{M}_P^2-\hat{M}_S^2}\, ,
\qquad\qquad
\hat{d}_m^2=\frac{F^2}{8}\frac{\hat{M}_S^2}{\hat{M}_P^2-\hat{M}_S^2}\,
\end{equation}
Though this has the same structure as the LO prediction~(\ref{eq.WSR-solution})
of the large--$N_C$ WSR in Eq.~(\ref{eq.largeNc-WSR}),
the couplings appearing here are the renormalized ones.  Nonetheless,
this result ensures that the difference between $\hat{c}_m/F$ and
$\hat{d}_m/F$ at NLO and their large--$N_C$ limits
remains small provided  $\hat{M}_R\approx M_R^{N_C\to\infty}$.
%%
%%Actually, it turns out to be zero if $\hat{M}_R=M_R^{N_C\to\infty}$.
%%
In order to achieve this minimization, the shifts $\Delta \kappa$
must be tuned in such a way that they obey
\begin{eqnarray}
  32 \hat{d}_m \Delta d_m -32 \hat{c}_m \Delta c_m\,\,+\,\,  A(\mu)
\,\,=\,\, 0  \,,
\nn\\
  32 \hat{M}_P^2 \hat{d}_m \Delta d_m  + 16 \hat{d}_m^2 \Delta M_P^2
-32 \hat{M}_S^2 \hat{c}_m \Delta c_m -  16 \hat{c}_m^2 \Delta M_S^2\,\,+\,\,  B(\mu)
\,\,=\,\, 0 \, .
\end{eqnarray}
If one fixes $\Delta M_R^2$ (for instance, through the pole scheme)
the solutions for $\Delta c_m$ and $\Delta  d_m$ are then given by
\begin{eqnarray}
32 \hat{c}_m\Delta c_m  &=&
\frac{  \hat{M}_P^2  A(\mu)   -B(\mu)+16\,\hat{c}_m^2
\Delta M_S^2-16\,\hat{d}_m^2\Delta M_P^2}
{ \hat{M}_P^2-\hat{M}_S^2 }
\nn\\
32 \hat{d}_m \Delta d_m &=&
\frac{\hat{M}_S^2 A(\mu) -  B(\mu)+16\,\hat{c}_m^2\Delta M_S^2
-  16 \, \hat{d}_m^2  \Delta M_P^2}
{ \hat{M}_P^2-\hat{M}_S^2 }\, .
\label{eq.WSR-scheme}
\end{eqnarray}

%%
%%This shifts will be relevant when we rewrite the amplitude
%%in the $\wms$--scheme into the other scheme.
%%
In the change of scheme we will make the replacement $\kappa=\hat{\kappa}+\Delta\kappa$
in the tree-level LO diagrams, whereas in the subleading contributions
we will just consider $\kappa\approx \hat{\kappa}$, as the difference goes
to NNLO in $1/N_C$. Thus, we will end up with a matrix element
expressed in terms of just renormalized couplings in the~new~scheme~($\hat{\kappa}$).
We will denote the $c_m$ and $d_m$ renormalization scheme
prescribed by Eq.~(\ref{eq.WSR-scheme}) as WSR--scheme.

\section{Low-energy expansion}
\label{sec.low}

\subsection{$\wms$--subtraction scheme}

At low energies, the expansion of our one-loop  \rcht\ correlator   yields
the   structure,
\begin{eqnarray}
\Pi_{ss-pp}(p^2) &=& B_0^2\,\,\left\{
\frac{2F^2}{p^2}
\,\,+\,\, \left[  \frac{16c_m^2}{M_S^2}-\frac{16d_m^2}{M_P^2}
+ 32\widetilde{L}_8 +
\frac{G_8}{\pi^2}\left(1-\ln\frac{-p^2}{\mu^2}\right)
+  32 \, \xi_{L_8} \right]\right.
\label{eq.lowRChT}\\
&&
\left.   + \Frac{p^2}{F^2} \left[
%%
%%32\widetilde{C}_{38} +
%%
\frac{16\, F^2 c_m^2}{M_S^4}-\frac{16\, F^2 d_m^2}{M_P^4}
-\Frac{G_{38}^L}{\pi^2}\left(1-\ln\frac{-p^2}{\mu^2}\right)+32 \,\xi_{C_{38}}
\,\,\, +\,\,\,\cO(N_C^0)\right] +{\cal O}(p^4)
\right\}\, ,
\nonumber
\end{eqnarray}
where in the $U(3)$ case we obtain
$G_8 = \frac{3}{16}= \Gamma_8$ and  $G_{38}^L= - 3c_dc_m/2 M_S^2$,
with  $G_{38}^L=\Gamma_{38}^L$ after using the LO matching relation
$L_5  = c_dc_m/  M_S^2$~\cite{Ecker1}.
The logarithm from the $\pi\pi$ loop in   \rcht\
has been singled out in the $\ln(-q^2)$ terms.
These \rcht\ logarithms exactly reproduce  those in the low-energy $\chi$PT
expression~(\ref{ChPT}),
ensuring the possibility of matching both theories~\cite{NLO-satura}.
The one-loop contributions from the remaining channels generate
only polynomial terms  at this chiral order and they  are provided here by
$\xi_{L_8}$ and $\xi_{C_{38}}$, defined within the $\wms$--renormalization-scheme.
The predictions for the low-energy constants at NLO in $1/N_C$
then turn out to be
\begin{eqnarray}
L_8(\mu_\chi) &=&  \frac{ c_m^2}{2\,M_S^2}-\frac{ d_m^2}{2\, M_P^2}
+  \widetilde{L}_8 +  \xi_{L_8}
\,\,\,\,\,+\,\Frac{\Gamma_8}{\pi^2}\ln\Frac{\mu^2}{\mu_\chi^2}\, ,
\nn\\
C_{38}(\mu_\chi)&=& \frac{ F^2 c_m^2}{2\, M_S^4}-\frac{ F^2 d_m^2}{2\,M_P^4}
+ \xi_{C_{38}}
\,\,\,\,\,- \, \Frac{\Gamma_{38}^L}{\pi^2}\ln\Frac{\mu^2}{\mu_\chi^2}
\, .
\end{eqnarray}
The dependence of the terms on the right-hand side  of the equations (r.h.s.)
on the \rcht\ renormalization scale $\mu$  have been left partially implicit.
Only the last term shows   $\mu$ explicitly.
It comes from the two--Goldstone loop in \rcht\ (Eq.~(\ref{eq.lowRChT}))
and matches  exactly the log from the two--Goldstone loop in \chpt\
(Eq.~(\ref{ChPT}),
with the chiral renormalization scale $\mu_\chi$),
producing   the  $\ln(\mu^2/\mu_\chi^2)$ term.
This ensures the right low-energy running with $\mu_\chi$
for the $\chi$PT low-energy constants~\cite{NLO-satura}.   On the other hand,
the r.h.s.  is independent of the \rcht\ scale $\mu$ at the given order in $1/N_C$.
There can still be some residual $\mu$ dependence at NNLO, which would allow
the use of renormalization group technics in order to improve the
perturbative expansion and to remove possible large radiative corrections~\cite{RGE}.
Nonetheless, this is beyond the scope of this article,
where we will take the usual prescription
$\mu=\mu_\chi$~\cite{L9,Rosell-thesis,NLO-satura}.

If we use the $c_m$ and $d_m$ predictions from the high-energy OPE
constraints in Eq.~(\ref{eq.NLO-WSR}), the low-energy predictions
result~\cite{L8}
\begin{eqnarray}
L_8(\mu)&=&  \Frac{F^2}{16}\,\left(\Frac{1}{M_S^2}+\Frac{1}{M_P^2}\right) \,
\left[1\,\,+\,\,\Frac{A(\mu)}{2 F^2}-\Frac{B(\mu)}{2 F^2 (M_S^2+M_P^2)}\right]
\,\,\,\, +\,\,\,\,\xi_{L_8}\, ,
\nn\\
\nn\\
C_{38}(\mu) &=& \Frac{F^4 \, (M_S^4+M_S^2 M_P^2 +M_P^4)}{
16 \, M_S^4 M_P^4 }\,
\left[ 1\,\,+\,\, \Frac{A(\mu)}{2 F^2} -\Frac{B(\mu) (M_S^2+M_P^2)}{
2 F^2 (M_S^4+M_S^2 M_P^2 +M_P^4)}\right]\,\,\,\, +\,\,\,\, \xi_{C_{38}}\, .
\nn\\
\label{eq.L8-linf}
\end{eqnarray}

In the case, where we only have interactions $\mL_R$ in the lagrangian,
linear in the resonance fields~\cite{Ecker1},
the low-energy contributions from the one-loop diagrams are given by
\begin{eqnarray}
\xi_{L_8} &=&    -
\frac{3c_dc_m}{32\pi^2F^2}\left(\ln\frac{M_S^2}{\mu^2}+\frac12
\right) +
\frac{3c_d^2}{128\pi^2F^2}\left(\ln\frac{M_S^2}{\mu^2}+\frac56\right)\nonumber\\&&
+\frac{3G_V^2}{256\pi^2F^2}\left(\ln\frac{M_V^2}{\mu^2}+\frac56\right)
+\frac{3c_m^2}{32\pi^2F^2}\ln\frac{M_S^2}{\mu^2}
-\frac{3d_m^2}{32\pi^2F^2}\ln\frac{M_P^2}{\mu^2}\, ,
\nn\\
\xi_{C_{38}}  &=&
 \frac{3d_m^2}{64\pi^2M_P^2}-\frac{3c_d^2}{512\pi^2M_S^2}
-\frac{3G_V^2}{1024\pi^2M_V^2}\nonumber\\&&-\frac{3c_m^2}{64\pi^2M_S^2}
+\frac{3c_dc_m}{96\pi^2M_S^2}
\, .
\end{eqnarray}

The results~(\ref{eq.L8-linf}) correspond to the predictions for the
$U(3)$ chiral perturbation theory couplings, where the $\eta_1$ is
identified as the ninth chiral Goldstone.  In order to recover the
traditional $SU(3)$ couplings one needs to make use of the matching
equations~\cite{L8,Leutwyler-Kaiser},
\begin{eqnarray}
L_8^{SU(3)}(\mu) &=& L_8^{U(3)}\,\,+\,\,
\Frac{\Gamma_8^{SU(3)}-\Gamma_8^{U(3)}}{32\pi^2}  \ln\Frac{m_0^2}{\mu^2} \, ,
\label{eq.SU3-match}\\
C_{38}^{SU(3)}(\mu) &=& C_{38}^{U(3)}\,\,- \,\,
\Frac{\Gamma_{38}^{(L)\, SU(3)}-\Gamma_{38}^{(L)\, U(3)}}{32\pi^2}
\,\left( \ln\Frac{m_0^2}{\mu^2} +\Frac{1}{2}\right)
\,-\, \Frac{\Gamma_8^{SU(3)}-\Gamma_8^{U(3)}}{32\pi^2} \,\, \Frac{F^2}{2 m_0^2}  \, .
\nn
\end{eqnarray}

These outcomes will be used later in the alternative renormalization schemes
and the constants $\xi_{L_8}(\mu)$, $\xi_{C_{38}}(\mu)$,
$A(\mu)$ and $B(\mu)$ will always refer to their former
expressions in the $\wms$ scheme.

\subsection{Pole masses and WSR--scheme for $c_m$ and $d_m$}

In this case, the renormalization scheme of $c_m$ and $d_m$ is
chosen such that the one-loop contributions to the NLO relations in
Eq.~(\ref{eq.NLO-WSR}) are exactly canceled, yielding
$\hat{c}_m^2=\frac{F^2}{8}\frac{\hat{M}_P^2}{\hat{M}_P^2-\hat{M}_S^2}$
and
$\hat{d}_m^2=\frac{F^2}{8}\frac{\hat{M}_S^2}{\hat{M}_P^2-\hat{M}_S^2}$.
The low energy limit of the \rcht\ correlator in the new scheme leads to
the LEC determination,
\begin{eqnarray}
L_8(\mu)&=&  \Frac{F^2}{16}\,\left(\Frac{1}{\hat{M}_S^2}+\Frac{1}{\hat{M}_P^2}\right) \,
\left[1\,\,+\,\,\Frac{A(\mu)}{2 F^2}-\Frac{B(\mu)}{2 F^2 (\hM_S^2+\hM_P^2)}\right]
\nn\\
\nn\\
&&\qquad \, - \, \Frac{F^2}{16}\, \left(\Frac{\Delta M_S^2}{\hM_S^4}+\Frac{\Delta M_P^2}{\hM_P^4}
\right)
\,\,\,\, +\,\,\,\,\xi_{L_8}\, ,
\nn\\
\nn\\
C_{38}(\mu) &=& \Frac{F^4 \, (\hat{M}_S^4+\hat{M}_S^2 \hat{M}_P^2 +\hat{M}_P^4)}{
16 \, \hM_S^4 \hM_P^4 }\,
\left[ 1\,\,+\,\, \Frac{A(\mu)}{2 F^2} -\Frac{B(\mu) (\hM_S^2+\hM_P^2)}{
2 F^2 (\hM_S^4+\hM_S^2 \hM_P^2 +\hM_P^4)}\right]
\nn\\
\nn\\
&&\qquad -\Frac{F^4}{16 \,\hM_S^2 \hM_P^2}
\,\left(\Frac{\Delta M_S^2 (\hM_S^2 + 2 \hM_P^2)}{\hM_S^4}
+\Frac{\Delta M_P^2 (2 \hM_S^2 +\hM_P^2)}{\hM_P^4}\right)\,\,\,\, +\,\,\,\, \xi_{C_{38}}\, ,
\end{eqnarray}
where $\xi_{L_8 , C_{38}}$ are the same one-loop contributions to
the LECs  computed before  in the $\wms$--subtraction scheme. The same
applies to $A(\mu)$ and $B(\mu)$, which were defined as the one-loop
contributions to the high-energy expansion coefficients in the
$\wms$--scheme.   In  $\xi_{L_8 }$,  $\xi_{C_{38}}$, $A(\mu)$ and
$B(\mu)$ we will use the couplings and masses in the new scheme
($\hc_m, \hd_m, \hM_R$) instead of the original ones in the
$\wms$--scheme ($c_m,d_m,M_R$), as the difference goes to NNLO in
$1/N_C$. The constants $\Delta M_R^2=M_R^2-\hM_R^2$ provide  the
difference between the mass $M_R$ in the $\wms$--scheme and its
value $\hM_R$ in another scheme. In this paper it will refer in
particular to the mass pole, although it accepts further
generalizations.

Notice that these expressions are similar to those in the
$\wms$--scheme~(\ref{eq.L8-linf}), up to the $\Delta M_R^2$ terms
that arise due to the change of mass prescription.
The WSR--scheme does not modify the low-energy prediction,
it just serves to reduce the uncertainties in the NLO Weinberg sum-rules.

Finally, in order to obtain the traditional $SU(3)$--\chpt\ LECs,
one should use again the matching Eq.~(\ref{eq.SU3-match}).

\section{Correlator with the extended \rcht\ lagrangian}
\label{sec.extra-operators}

Ecker {\it et al.}'s lagrangian~\cite{Ecker1} has been found to be
very successful for the description of amplitudes with
few-Goldstones ($\pi\pi$ form-factors,  scatterings...). However, it
fails to describe processes with multi-Goldstones states or with a
higher number of resonances. The LO meson lagrangian must be then
enlarged to improve the description of the new
channels. In the case of our observable, the relevant
operators with two resonance
fields are~\cite{rcht-op6,L10,Rosell-genera,Rosell-thesis},
\begin{equation}
\mL_{RR'}\,=\, i \lambda_1^{PV} \bra [\nabla^\mu P,V_{\mu\nu}]\, u^\nu\ket
\,+\,   \lambda_1^{SA} \bra \{\nabla^\mu S,A_{\mu\nu}\}\, u^\nu\ket
\,+\,   \lambda_1^{SP} \bra \{ \nabla^\mu S,P\} \, u_\mu\ket\, .
\label{eq.lambda-lagr}
\end{equation}

The $\lambda_1^{PV}$ and $\lambda_1^{SP}$ terms induce a
one-loop mixing between the Goldstone and the pseudoscalar
resonance.  These loops bring ultraviolet divergences which need the
presence of the subleading counter-terms,
\begin{equation}
\Delta \mL_{P}\,=\, d_m'  \bra P \nabla_\mu u^\mu\ket
\,+\,   d_m'' \bra (\nabla^2 P)  \nabla_\mu u^\mu\ket \, ,
\end{equation}
to make the amplitude finite.  At LO, in the free field case,
the meson kinetic terms are assumed to be defined in the canonical way, i.e.,
without mixing between particles. This was indeed the case in Ecker {\it et al.}'s
lagrangian~\cite{Ecker1}.  In addition, although
these $P$--$\phi$ operators may arise at NLO, they happen to be proportional
to the EOM. They can be removed from the action through a convenient meson field
redefinition, leaving for the relevant couplings in our problem the
effective combinations
\begin{eqnarray}
\widetilde{L}_8^{eff} &=& \widetilde{L}_8 +\frac{1}{2}c_m^2X_S-\frac{1}{2}d_m^2X_P +
c_m\lambda^S_{18} - d_m\lambda^P_{13}   - \frac{1}{2} d_m d_m''\,   ,
\nonumber\\
\widetilde{H}_2^{eff} &=& \widetilde{H}_2 +c_m^2X_S+d_m^2X_P +
2c_m\lambda^S_{18} + 2d_m\lambda^P_{13} + d_m d_m''     \, ,
\nonumber\\
(M_S^2)^{eff} &=& M_S^2 - X_SM_S^4,\nonumber\\
(M_P^2)^{eff} &=& M_P^2 - X_PM_P^4,\nonumber\\
%%
%%c_d^{eff} &=& c_d - c_dX_SM_S^2\nonumber,\\
%%
c_m^{eff} &=& c_m - c_mX_SM_S^2 -M_S^2\lambda^S_{18},\nonumber\\
d_m^{eff} &=& d_m - d_mX_PM_P^2 - M_P^2\lambda^P_{13} + \frac{1}{2} d_m'
-\frac{1}{2} M_P^2 d_m''   \, .
\label{eq.eff-couplings2}
\end{eqnarray}

\subsection{Meson self-energies}

\begin{figure}[t!]
\begin{center}
\epsfxsize=8cm\epsfbox{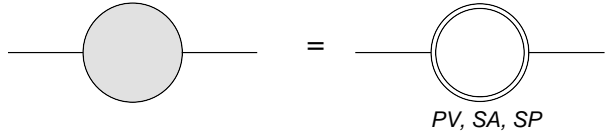}
\end{center}
\caption{{\small Contribution from $\mL_{RR'}$ operators
to the  Goldstone boson self-energy.}}
\label{fig.propG}
\end{figure}

These operators do not modify the previous loop contributions.
However, new channels are now open in the different
vertex-functions. Thus, the Goldstone self-energy gains the
contributions (Fig.~\ref{fig.propG}),
\begin{eqnarray}
 \Sigma^r_\phi(p^2)|_{PV} &=&
-\frac{3 (\lambda^{PV}_1)^2}{F^2 }\Bigg\{\left[M_V^4-2M_V^2(p^2+M_P^2)+(p^2-M_P^2)^2\right]\bar{J}(p^2,M_P^2,M_V^2)\nonumber\\&&
-\frac{p^2}{16\pi^2(M_P^2-M_V^2)}\left(p^2M_P^2\ln\frac{M_P^2}{\mu^2}-p^2M_V^2\ln\frac{M_V^2}{\mu^2}
+M_V^2M_P^2\ln\frac{M_V^2}{M_P^2}\right)-\frac{p^2(M_P^2+M_V^2)}{32\pi^2}\Bigg\} \, ,
\nonumber\\
 \Sigma^r_\phi(p^2)|_{SA}
&=&
-\frac{3 (\lambda^{SA}_1)^2}{F^2 }\Bigg\{\left[M_A^4-2M_A^2(p^2+M_S^2)+(p^2-M_S^2)^2\right]\bar{J}(p^2,M_S^2,M_A^2)\nonumber\\&&
-\frac{p^2}{16\pi^2(M_S^2-M_A^2)}\left(p^2M_S^2\ln\frac{M_S^2}{\mu^2}-p^2M_A^2\ln\frac{M_A^2}{\mu^2}
+M_A^2M_S^2\ln\frac{M_A^2}{M_S^2}\right)-\frac{p^2(M_S^2+M_A^2)}{32\pi^2}\Bigg\} \, ,
\nonumber\\
 \Sigma_\phi^r(p^2)|_{SP}&=&
-\frac{3 (\lambda^{SP}_1)^2}{p^4F^2}\Bigg\{(p^2-M_P^2+M_S^2)^2\bar{J}(p^2,M_P^2,M_S^2)\nonumber\\&&
-\frac{p^2}{16\pi^2(M_P^2-M_S^2)}\left(p^2M_P^2\ln\frac{M_P^2}{\mu^2}-p^2M_S^2\ln\frac{M_S^2}{\mu^2}
+M_S^2M_P^2\ln\frac{M_S^2}{M_P^2}\right)-\frac{p^2(M_P^2+M_S^2)}{32\pi^2}\Bigg\}\, ,
\nn\\
\end{eqnarray}
in addition to the former $V\phi$ and $S\phi$ cuts from Eq.~(\ref{eq.pion-self1}).
The functions $\bar{J}(p^2,M_a^2,M_b^2)$ is the subtracted
two-propagator Feynman integral  ($\bar{J}(0,M_a^2,M_b^2)=0$),
given in App.~\ref{app.Feynman}.

\begin{figure}[t!]
\begin{center}
\epsfxsize=7cm\epsfbox{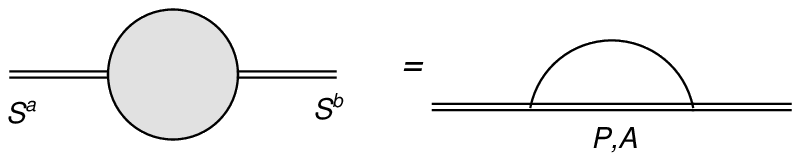} \hspace*{1.5cm}
\epsfxsize=7cm\epsfbox{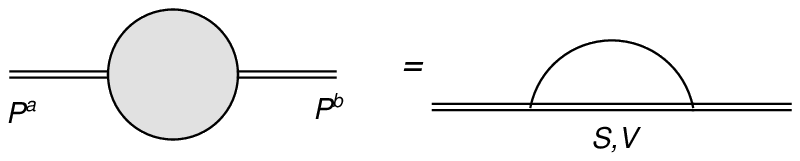}
\end{center}
\caption{{\small Contribution from $\mL_{RR'}$ operators to the
scalar and the pseudo-scalar resonance self-energies.}} \label{fig.propPS}
\end{figure}

The scalar propagators contains now $A\phi$ and $P\phi$ cuts
(besides the $\phi\phi$--one from Eq.~(\ref{eq.S-self1}))
(Fig.~\ref{fig.propPS}):
\begin{eqnarray}
\Sigma_S^r(p^2)|_{A\phi} &=&
\frac{3(\lambda^{SA}_1)^2}{16\pi^2F^2}\left[
\frac{(p^2-M_A^2)^3}{p^2}\ln\left(1-\frac{p^2}{M_A^2}\right) -M_A^4
-3p^2M_A^2\left(\ln\frac{M_A^2}{\mu^2}-\frac23\right)
+p^4\left(\ln\frac{M_A^2}{\mu^2}-1\right)\right]  ,
\nonumber\\
\Sigma_S^r(p^2)|_{P\phi} &=&
 \frac{3(\lambda^{SP}_1)^2}{16\pi^2F^2}\left[
\frac{(p^2-M_P^2)^3}{p^2}\ln\left(1-\frac{p^2}{M_P^2}\right) -M_P^4
+p^2M_P^2\left(\ln\frac{M_P^2}{\mu^2}+2\right) +
p^4\left(\frac{M_P^2}{\mu^2}-1\right) \right]   .
\nn\\
\end{eqnarray}

Ecker {\it et al.}'s lagrangian $\mL_R$ did not modified the
pseudoscalar resonance propagator. However, the new operators
$\mL_{RR'}$ yield
(Fig.~\ref{fig.propPS}),
\begin{eqnarray}
\Sigma_P^r(p^2)|_{V\phi} &=&
\frac{3(\lambda^{PV}_1)^2}{16\pi^2F^2}\left[
\frac{(p^2-M_V^2)^3}{p^2}\ln\left(1-\frac{p^2}{M_V^2}\right)
-M_V^4
\right.\nonumber\\ &&\hspace{3cm}\left.
-3p^2M_V^2\left(\ln\frac{M_V^2}{\mu^2}-\frac23\right)
+p^4\left(\ln\frac{M_V^2}{\mu^2}-1\right)
\right]  ,
\nonumber\\
\Sigma_P^r(p^2)|_{S\phi}  &=&
\frac{3(\lambda^{SP}_1)^2}{16\pi^2F^2}\left[
\frac{(p^2-M_S^2)^3}{p^2}\ln\left(1-\frac{p^2}{M_S^2}\right)
+4M_S^4\left(\ln\frac{M_S^2}{\mu^2}-\frac14\right)
\right.\nonumber \\&&\hspace{3cm}\left.
-3p^2M_S^2\left(\ln\frac{M_S^2}{\mu^2}-\frac23\right)
+p^4\left(\ln\frac{M_S^2}{\mu^2}-1\right) \right]  .
\nn\\
\end{eqnarray}

The renormalized resonance self-energies provide at this order the
pole masses through Eq.~(\ref{eq.pole-pos}), giving the
corresponding shifts $\Delta M_R^2=-$Re$\Sigma_R^r(M_R^2)$.

\subsection{$P$--$\phi$ mixing}

\begin{figure}[t!]
\begin{center}
\epsfxsize=7cm\epsfbox{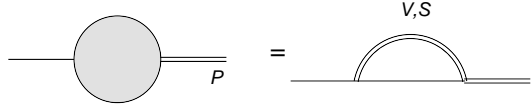}
\end{center}
\caption{{\small Contribution from $\mL_{RR'}$ operators to the
mixing term between the Goldstone and the pseudo-scalar resonance.}}
\label{fig.mix}
\end{figure}

In addition, these operators $\lambda_1^{PV}$ and $\lambda_1^{SP}$
also generate a $P$--$\phi$  mixing  (Fig.~\ref{fig.mix}),
\begin{eqnarray}
\Sigma_{P-\phi}^r(p^2)|_{V\phi} &=&
\frac{3G_V\lambda^{PV}_1}{16\pi^2F^3}\left[\frac{(p^2-M_V^2)^3}{p^2}\ln\left(1-\frac{p^2}{M_V^2}\right)
-M_V^4
\right.\nonumber \\&&\hspace{3cm}\left.
-3p^2M_V^2\left(\ln\frac{M_V^2}{\mu^2}-\frac23\right)
+p^4\left(\ln\frac{M_V^2}{\mu^2}-1\right)\right] ,
\nonumber\\
\Sigma_{P-\phi}^r(p^2)|_{S\phi}  &=&
\frac{3c_d\lambda^{SP}_1}{8\sqrt2\pi^2F^3}\left[\frac{(p^2-M_S^2)^3}{p^2}\ln\left(1-\frac{p^2}{M_S^2}\right)
-M_S^4
\right.\nonumber \\&&\hspace{3cm}\left.
-3p^2M_S^2\left(\ln\frac{M_S^2}{\mu^2}-\frac23\right)
+p^4\left(\ln\frac{M_S^2}{\mu^2}-1\right)\right] ,
\end{eqnarray}
leading to an extra  perturbative contribution to the $PP$--correlator
that has to be added to the former ones in~(\ref{eq.general-PI}):
\begin{eqnarray}
\Pi_{pp}(p^2)^{\rm P-\phi\,  \, mixing} &=&
\Frac{ 8\sqrt{2}\, F  d_m}{p^2 \, (p^2-M_P^2)}\,\, \left[\,
-\frac{\sqrt{2} \, d_m'}{F} p^2 +\frac{\sqrt{2} \,d_m''}{F}p^4
\,+\, \Sigma_{P-\phi}^r(p^2)^{1\ell}\, \right]\, .
\label{eq.PI-mixing}
\end{eqnarray}
After a convenient  field redefinition $d_m'$ and $d_m''$ disappear
from Eq.~(\ref{eq.PI-mixing}), being their information encoded in
$d_m^{\rm eff}$, $\widetilde{L}_8^{\rm eff}$ and
$\widetilde{H}_2^{\rm eff}$.

It is important to remark that at the NLO under consideration,
the mixing does not modify the pseudoscalar resonance mass renormalization.
The Goldstone remains massless --as expected-- and the
resonance pole mass
is still provided at this order by  $\Delta M_P^2=-$Re$\Sigma_P^r(M_P^2)$
through Eq.~(\ref{eq.pole-pos}).

\subsection{New $s\to S$ and $p\to P$ vertex functions}

A new $P\phi$ channel is opened in the $s\to S$ vertex function in addition to
the $\phi\phi$--cut from Eq.~(\ref{eq.sS-vertex1}):
\begin{eqnarray}
-\Frac{1}{ 4 B_0} \Phi^r_{sS}(p^2)^{1\ell} |_{P\phi} &=&
\frac{3d_m\lambda^{SP}_1}{16\pi^2F^2}\left[\frac{(p^2-M_P^2)^2}{p^2}\ln\left(1-\frac{p^2}{M_P^2}\right)
+M_P^2+p^2\left(\ln\frac{M_P^2}{\mu^2}-1\right)\right]
.
\label{eq.sS-vertex2}
\end{eqnarray}
On the other hand, one has now the $S\phi$--absorptive cut
in the $p\to P$ vertex-function, which did not get any contribution
from $\mL_R$ alone:
\begin{eqnarray}
-\Frac{1}{ 4 B_0} \Phi^r_{pP}(p^2)^{1\ell} |_{S\phi} &=&
\frac{3c_m\lambda^{SP}_1}{16\pi^2F^2}\left[\frac{(p^2-M_S^2)^2}{p^2}\ln\left(1-\frac{p^2}{M_S^2}\right)
\right.\nonumber \\&&\hspace{3cm}\left.
-2M_S^2\left(\ln\frac{M_S^2}{\mu^2}-\frac12\right)+p^2\left(\ln\frac{M_S^2}{\mu^2}-1\right)\right].
\label{eq.sS-vertex2}
\end{eqnarray}

\begin{figure}[t!]
\begin{center}
\epsfxsize=6cm\epsfbox{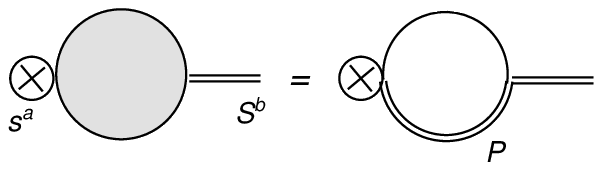} \hspace*{1.5cm}
\epsfxsize=6cm\epsfbox{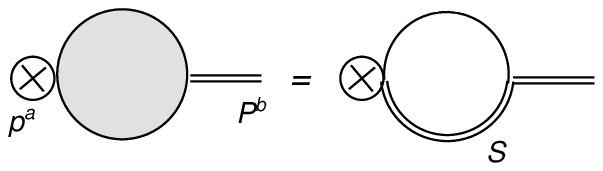}
\end{center}
\caption{{\small One-loop diagrams  with  $\mL_{RR'}$ operators in
the $s(x)\to S$ and $p(x)\to P$ vertex functions.}} \label{fig.vert}
\end{figure}

\section{Phenomenology}
\label{sec.pheno}

The \rcht\ lagrangian developed by Ecker {\it et al.}~\cite{Ecker1},
$\mL=\mL_{G}+\mL_R$, only contained operators with at most one
resonance field. This approach has been proven to be very successful
at the phenomenological level for the last two decades~\cite{PI:08}.
Nevertheless, in the few last years it has become clear that the
description of more complicated QCD matrix elements (e.g. 3--point
Green functions~\cite{Pich1,rcht-op6,Pich3,Pich4,Prades-GF})
demands the introduction of
operators with more than one resonance field~\cite{rcht-op6}.

Since the $M_R(\mu)$ masses in the $\wms$---scheme are $\mu$
dependent, they are difficult to relate with the physical masses
provided, for instance,  by the {\it  Particle Data Group}
(PDG)~\cite{PDG}. This relation is even more cumbersome when
one adds more general kinds of vertices (e.g. $\lambda_1^{SP}$)
within the loops:  in the $\wms$--scheme  the value of $M_R(\mu)$ will
depend on the content of the theory and its lagrangian. Thus, it
seems more convenient to use  universal properties such as the pole masses,
denoted here as  $\hM_R$. The octet of the lightest scalar and the
pseudoscalar resonances are then related, to the $a_0(980)$ and the
$\pi(1300)$, and we will consider from now on the inputs $\hM_S=980
\pm 20$~MeV and $\hM_P=1300\pm 50$~MeV~\cite{Guo-aIJ,PDG}.

The procedure that we will follow in order to extract the LECs with higher and higher
accuracy is to progressively add more and more physical information
to the \rcht\ correlator,  starting from lower
energies.   Since the resonance parameters will be used
to accommodate the short-distance OPE behaviour,
in general the two-meson thresholds ($S\pi$, $V\pi$, $P\pi$...)
may not be at the right place.   Likewise, one may
find that individual intermediate two-meson channels
have a clearly erroneous momentum  dependence at high energies (e.g.
constant or growing  behaviour).

The introduction of the new operators $\lambda_1^{VP}$,
$\lambda_1^{SP}$ and $\lambda_1^{SA}$
will allow us to improve the  momentum dependence of the $R\pi$ absorptive
channels with one resonance and one Goldstone.
However, since these new couplings will be tuned to
implement the short-distance OPE constraints,
the  $R\pi$ channel description
may still differ slightly  from that provided by the physical values of
$\lambda_{SP}$, $\lambda_{PV}$, $\lambda_{SA}$, $c_d$, $G_V$...
Likewise,   the two-resonance $RR'$ absorptive cuts will still remain
wrongly described until operators with three resonance fields are taken into
account.
Nonetheless, we will see that
the \rcht\ description progressively approaches the
actual QCD amplitude as the hadronic action is completed with more and more
complicated operators, bringing along a better and better
description of the lower channels.

\subsection{Phenomenology with Ecker {\it et al.}'s lagrangian
$\mL_G+\mL_R$}

First, we will extract the value of the LECs at large $N_C$ within
the single resonance approximation.  We will use the formerly
referred $\hM_S=980 \pm 20$~MeV and $\hM_P=1300\pm
50$~MeV~\cite{PDG}, $F=90\pm 2$~MeV~\cite{Guo-aIJ,Leutwyler-Kaiser}
and the standard reference $\chi$PT renormalization scale
$\mu_0=770$~MeV. The short-distance constraints determine
$c_m$ and $d_m$ in terms of the scalar and pseudo-scalar masses, producing
\begin{equation}
L_8\,=\, (0.83\pm 0.05)\, \cdot 10^{-3}\,, \qquad\qquad
C_{38}\,= \, \,(8.4\pm 1.0) \cdot\, 10^{-6}\, .
\end{equation}
Naively, if  the uncertainty on the saturation scale is estimated  by
observing the variation with $\mu$ in the range 0.5--1~GeV, one would expect
the former values to be deviated from the actual ones at the order
of $\Delta L_8\sim  0.3 \cdot 10^{-3}$, $\Delta C_{38}\sim   5\cdot 10^{-6}$.

In order to go beyond the naive estimate of the subleading $1/N_C$
uncertainty, we consider now the one-loop contributions computed in
previous sections. In a first approach, we  consider just operators
in the lagrangian with at most one resonance field~\cite{Ecker1}.
At one-loop,
in addition to the tree-level exchanges, one has the two-meson
absorptive channels $\pi\pi$, $V\pi$, $S\pi$ and $P\pi$, determined
by the scalar parameters $c_m$ and $c_d$, the pseudo-scalar coupling
$d_m$ and the vector ones $G_V$ and $M_V$. If we work in the
WSR--renormalization-scheme for $c_m$ and $d_m$, the short-distance
constraints  produce at NLO the same structure
found from the large--$N_C$ WSR, $\hat{c}_m^2=\frac{F^2}{8}\frac{\hM_P^2}{\hM_P^2-\hM_S^2}$
and $\hat{d}_m^2=\frac{F^2}{8}\frac{\hM_S^2}{\hM_P^2-\hM_S^2}$. The
other three resonance parameters ($c_d,G_V,M_V$) are fixed by means
of the logarithmic OPE constraints~(\ref{eq.Ecker-log-const}),
$\alpha_0^{(\ell)}=\alpha_2^{(\ell)}=\alpha_4^{(\ell)}=0$,  giving
\begin{equation}
c_d\,=\, 60\,\pm \, 4\, \mbox{MeV}\, , \qquad\qquad
G_V\,=\, 93\,\pm \, 5\, \mbox{MeV}\, , \qquad\qquad
M_V\,=\,853\,\pm \, 28\, \mbox{MeV}\, .
\end{equation}
These numbers are found to be quite off the physical
ones,
$c_d\approx 30$~MeV, $G_V\approx 60$~MeV,
$M_V\approx
770$~MeV~\cite{Ecker1,Ecker2,Ivashyn1,Ivashyn2,Jamin-cdcm,Masjuan-cd,Guo-aIJ,PDG}.
The LEC prediction  for the standard comparison scale ${ \mu_0=770 }$~MeV then result,
\begin{equation}
L_8(\mu_0)\,= \, (2.28 \,\pm\, 0.19)\, \cdot 10^{-3}\,, \qquad\qquad
C_{38}(\mu_0)\,= \,(26\, \pm\, 4 ) \, \cdot\, 10^{-6}\, .
\end{equation}
In order to get these $SU(3)$ \chpt\ couplings, we employed
in the U(3)--SU(3) matching Eq.~(\ref{eq.SU3-match})
the chiral singlet pseudoscalar mass $m_0=850\pm 50$~MeV~\cite{Leutwyler-Kaiser}.
These estimates are still far from former  values in the bibliography
for $\mu_0=770$~MeV:
$L_8=0.9 \cdot 10^{-3}$ and $C_{38}=10\cdot 10^{-6}$ from
$\cO(p^6)$ \chpt\ and resonance estimates~\cite{SS-PP-Bijnens},
later refined into  $L_8=(0.61\pm 0.20)\cdot 10^{-3}$~\cite{fit10}
and recently updated into $L_8=(0.37\pm 0.17)\cdot 10^{-3}$~\cite{new-fit10};
$L_8=(0.6\pm 0.4) \cdot 10^{-3}$ and $C_{38}=(2\pm 6) \cdot 10^{-6}$ from
a previous NLO calculation in R$\chi$T~\cite{L8};
$L_8=(1.02 \pm 0.06)  \cdot 10^{-3}$ and
$C_{38}=(3.3\pm 0.6) \cdot 10^{-6}$ from Dyson-Schwinger equation
analysis~\cite{DSE};
$L_8 = (0.36\pm 0.05\pm 0.07)\cdot 10^{-3}$ from Lattice simulations~\cite{MILC}.

Although the calculation with just the $\mL_R$ operators is able
produce an appropriate description of the $\pi\pi$ channel (thanks
to the $c_d\bra S u_\mu u^\mu\ket$ operator),  its coupling $c_d$
gets a extremely shifted value as this parameter  has been used to
accommodate the OPE at  short distances. This does not represent by
itself an important drawback in our analysis, where the goals are
the LECs and \rcht\ is devised as a convenient interpolator between
high and low energies. However, the problem in our case is the
erroneous description that one obtains for the $R\pi$ channels with
only the $\mL_R$ operators~\cite{L10,Rosell-thesis}: The $S\pi$
contribution to the spectral function behaves like a constant and
the $V\pi$ one grows with the energy. moreover,  as $M_V$ is also
determined from the OPE matching, the position of the first
two-meson threshold after the $\pi\pi$ one (i.e., the $V\pi$
channel) is shifted from its physical place.

\subsection{Improving one $R\pi$ channel: extending the lagrangian}

The straight forward procedure to ameliorate our one-loop amplitude
is the inclusion of the required operators for the proper
description of the lowest absorptive cuts,  this is, $\pi\pi$ and $V\pi$. The
first one is ruled by the already included $c_d$ operator but the
latter demands the $\lambda_1^{PV}$ term  from
Eq.~(\ref{eq.lambda-lagr}), which now induces   $PV\pi$ interactions
and allows to cure the infinitely growing behaviour of the $V\pi$
contribution to the spectral function. %%
%%For this the couplings needs to take the value
%%$\lambda_1^{PV}= G_V/2\sqrt{2} d_m$~\cite{L10,Rosell-thesis},
%%which numerically  turns out to be $|\lambda_1^{PV}|\approx 0.7$~\cite{PI:08}.
%%

Now we use the former inputs $\hM_S$, $\hM_P$, $F$, $m_0$    and
the physical coupling $c_d=30\pm
10$~MeV~\cite{Ecker1,Jamin-cdcm,Masjuan-cd,Guo-aIJ,PI:08}. The
remaining parameters ($G_V, M_V,\lambda_1^{PV}$) are extracted from
the three logarithmic OPE constraints
$\alpha_0^{(\ell)}=\alpha_2^{(\ell)}=\alpha_4^{(\ell)}=0$. Indeed,
this system only has real solutions in the very corner of the
parameter space, for low pseudo-scalar  mass ($\hM_P\approx
1.25$~GeV) and high $c_d$ and scalar mass ($c_d\approx 40$~MeV,
$\hM_S\approx 1.00$~GeV). This does not improve the value of the
vector coupling and mass with respect to the former section, which
become $G_V\approx  120$~MeV and $M_V\approx 400$~MeV. The LEC
predictions result,
\begin{equation}
L_8(\mu_0)\,\approx \, 0.5 \, \cdot \, 10^{-3}\, ,\qquad\qquad
C_{38}(\mu_0)\,\approx \, -8 \,\cdot\, 10^{-6}\, ,
\end{equation}
where $L_8$ may look acceptable but the presence of such a low
distorted $V\pi$ threshold is reflected in a value of $C_{38}$
which looks still a bit off.  Nonetheless,
these values are closer to those formerly obtained
in the bibliography~\cite{SS-PP-Bijnens,fit10,new-fit10,L8,DSE,MILC}.

The problem is that the $V\pi$ is not the only relevant channel that
appears after the $\pi\pi$ one. The $S\pi$ channel opens  up at an
energy not far from the $V\pi$ threshold.  Thus, even if  the $V\pi$
channel can be now correctly described, the  $S\pi$ contribution to
the spectral function still shows a wrong constant
behaviour~\cite{L10,Rosell-thesis}.   The $\lambda_1^{SP}$ operator
in~(\ref{eq.lambda-lagr}) is then crucial to cure that behaviour. %%
%%, which
%%should need the value $\lambda_1^{SP}=(c_d-2 c_m)/2 d_m$~\cite{L10,Rosell-thesis},
%%numerically $|\lambda_1^{SP}|\approx 0.7$~\cite{PI:08}.
%%
Furthermore, this operator  mends as well  the similar bad short-distance behaviour
found in  the $P\pi$ cut contribution  to the $SS$ spectral function.

Nonetheless,  the presence of $\lambda_1^{PV}$ in the lagrangian is still
essential.  If one repeats the NLO computation adding only the $\lambda_1^{SP}$
operator  (but not $\lambda_1^{PV}$)
the vector parameters become of the order of
$G_V\sim 20$~MeV and $M_V\sim 2$~GeV. On the other hand, the LEC predictions
$L_8\sim 1.3\cdot 10^{-3} $  and
$C_{38}\sim 12\cdot 10^{-6}$  seem to improve with respect to the case with
only $\mL_R$ operators in the \rcht\ lagrangian~\cite{Ecker1},
with at most one resonance field.

The inclusion of the $\lambda_1^{SA}$ operator  alone
seems to move the results also in the right direction.  Although it does not affect
the previous channels, it opens the $A\pi$ absorptive cut.
Even if its effect at low energies is   small, it helps to fulfill
the OPE constraints. Taking now the extra needed input $M_V=770\pm 20$~MeV
together with the former ones,  it is possible to extract the remaining
ones $(\lambda_1^{SA}, G_V,M_A)$ through the three log OPE conditions.
The value for the vector coupling turns out to be now more natural
($G_V=67\pm18$~MeV) but the $a_1(1230)$ mass falls down  to very low values
($M_A=610\pm 50$~MeV).
The predictions for the chiral couplings show a clear improvement,
$ L_8=(0.7\pm0.4)\cdot 10^{-3}$,
$C_{38}=(4\pm 5)\cdot 10^{-6}$.

\subsection{Improving the $V\pi$, $S\pi$, $A\pi$ and $P\pi$   channels}
%%%
%%%Analysis with  $\lambda_1^{PV}$ and $\lambda_1^{SP}$}
%%%
%%%If one now includes the  operators   $\lambda_1^{PV}$ and $\lambda_1^{SP}$  from
%%%Eq.~(\ref{eq.lambda-lagr}) in the \rcht\ lagrangian,
%%%the situation shows a clear improvement.  Together with the former inputs
%%%$\hM_S$, $\hM_P$, $F$, $m_0$,  we now add the $S\pi\pi$ coupling
%%%$c_d=30\pm 10$~MeV and the right $V\pi$ threshold position
%%%$M_V=770\pm 20$~MeV. The remaining parameters ($G_V,\lambda_1^{PV}, \lambda_1^{SP}$)
%%%are fixed through the three logarithmic OPE constraints.
%%%The vector coupling is produced  then in the range $G_V\sim 50$--$60$~MeV,
%%%close enough to its physical value,  getting now a  clear improvement
%%%in the LEC predictions,
%%%\begin{equation}
%%%L_8(\mu_0)\,\approx \, (0.9 \pm) \, \cdot \, 10^{-3}\, ,\qquad\qquad
%%%C_{38}(\mu_0)\,\approx \, (8 \pm )\,\cdot\, 10^{-6}\, .
%%%\end{equation}
%%%

\begin{figure}[t!]
\begin{center}
\includegraphics[angle=0,width=8cm,clip]{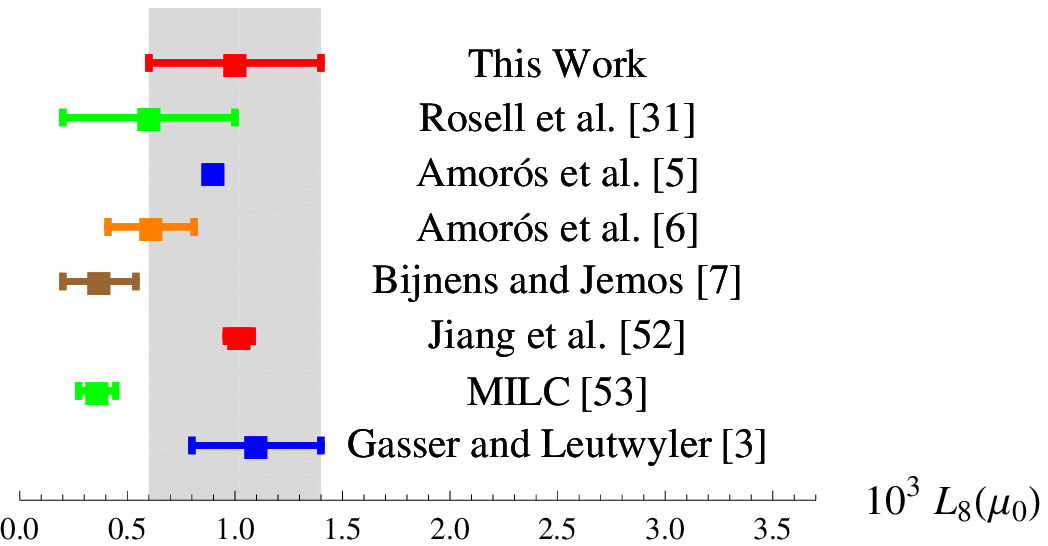}
\hspace*{0.5cm}
\includegraphics[angle=0,width=8cm,clip]{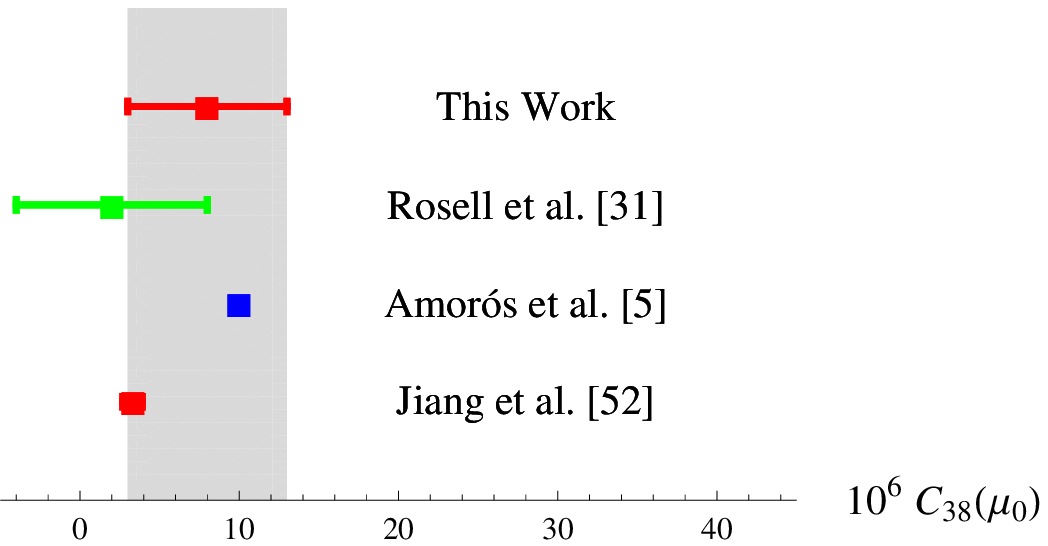}
\end{center}
\caption{{\small
Comparison of the LEC predictions in this work with previous results in the
bibliography.  }}
\label{fig.LECs}
\end{figure}

In order to have a proper description of all the $R\pi$ absorptive cuts,
the $\lambda_1^{SA}$, $\lambda_1^{PV}$ and $\lambda_1^{SP}$
operators from Eq.~(\ref{eq.lambda-lagr}) are now included in the
\rcht\ action.
We take the same inputs as before,
$\hM_S=980\pm 20$~MeV, $\hM_P=1300\pm50$~MeV, $F=90\pm 2$~MeV,
$m_0=850\pm 50$~MeV,  $c_d=30\pm 10$~MeV, $M_A=1230\pm 200$~MeV,
$M_V=770\pm 20$~MeV and $G_V=60\pm 20$~MeV.
Both $c_d$ and $G_V$ have been taken with a naive 33\%
error, as they appear only in the NLO part of the correlator.  This will
account for the possible NNLO variations in the one-loop correlator
depending on whether it is evaluated with these physical couplings
or their large--$N_C$ values.
The remaining unknown parameters  ($\lambda_1^{PV}, \lambda_1^{SP},\lambda_1^{SA}$)
are extracted from the three logarithmic OPE constraints,
leading to our final LEC estimates,
\begin{equation}
L_8(\mu_0)\,= \, (1.0 \pm 0.4) \, \cdot \, 10^{-3}\, ,\qquad\qquad
C_{38}(\mu_0)\, = \, (8 \pm 5)\,\cdot\, 10^{-6}\, .
\label{eq.final-LEC}
\end{equation}
These numbers are compared to previous determinations in Fig.~\ref{fig.LECs}.
Although there is still a clear dispersion between the various measurements,
at the present error level we remain essentially compatible.
Further efforts should be focused on the extraction of the scalar and
pseudo-scalar pole masses in order to sizably reduce the uncertainties
in the \rcht\ calculations.

\begin{figure}[t!]
\begin{center}
\includegraphics[angle=0,width=5.5cm,clip]{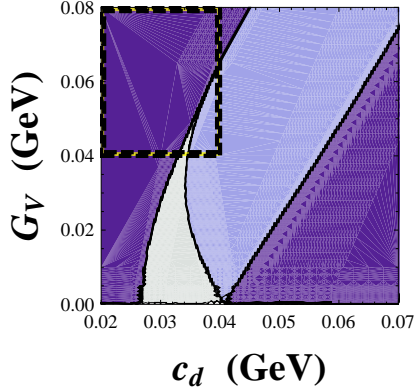}
\end{center}
\caption{{\small Allowed regions with one (light blue) or two (white) real
solutions from
the logarithmic OPE constraints
$\alpha_0^{(\ell)}=\alpha_2^{(\ell)}=\alpha_4^{(\ell)}=0$.
No real solution exists in the darker purple
regions. In the upper left corner one can see the dashed rectangle
provided by the ranges $c_d=30\pm 10$~MeV, $G_V=60\pm 20$~MeV.  }}
\label{fig.allowed}
\end{figure}

In general, the three logarithmic OPE constraints
$\alpha_0^{(\ell)}=\alpha_2^{(\ell)}=\alpha_4^{(\ell)}=0$ produce
complex solutions for the $\lambda_1^{SP}$, $\lambda_1^{PV}$,
$\lambda_1^{SA}$. In order to remain within the  quantum field
theory description, only the real values are kept. The regions with
at least one real solution are shown in Fig.~\ref{fig.allowed}.
There, we plot the allowed ranges for $c_d$ and $G_V$, with the
other inputs taken at their central values. Indeed, there is no real
solution for the central values $c_d=30$~MeV and $G_V=60$~MeV.  On
the contrary  to other phenomenological analysis which seem to
prefer a $c_d$ coupling below 30~MeV~\cite{Ivashyn1,Masjuan-cd,Guo-aIJ}, the
log OPE constraints require slightly larger values, $cd\gsim
30$~MeV. However, in general for $c_d$ around $30$~MeV is always
impossible to have real solutions for the value of the coupling
$G_V\simeq 64$~MeV obtained from $V$ decays~\cite{Ecker1,Ecker2,Guo-aIJ}.
Actually, if one demanded the $\pi\pi$ scalar form-factor (and the
corresponding $\pi\pi$ contribution to the $SS$ spectral function)
to vanish at high energies one would obtain $c_d=F^2/4 c_m\simeq
42$~MeV. However, in this work we do not perform a channel by
channel analysis as in Ref.~\cite{L8}. Indeed, in our field theory
approach one could fix separately the short-distance behaviour of the $\pi\pi$
and all the $R\pi$ channels through the $\lambda^{RR'}$ operators,
but the latter also generate $RR'$ absorptive cuts with the wrong properties
at high momentum.   The only option is the global adjustment  of
parameters considered in this work, where the lowest channels
arrange the short-distance behaviour of the highest cuts at the
price of slight modifications on their couplings.

The allowed $(c_d, G_V)$ region of Fig.~\ref{fig.allowed} actually
changes if one varies the other inputs.  Thus, we observed the whole
range of the LECs allowed for the possible variations of the inputs
and used this interval as our estimate of the central value and
error. The maximum (minimum)  value of the LECs was obtained at the
largest (smallest) $c_d$ and $G_V$.  Likewise, the most extreme LEC
values were obtain when $\hM_P$ and $M_A$ became smaller and $\hM_S$
larger. These three parameters are responsible for most of the
uncertainties. The impact of the $M_V$, $F$ and $m_0$ errors in the
global precision is negligible.

The \rcht\ computation progressively approaches the physical  value as
one incorporates more and more physical information. This is quite
non-trivial,  as the introduction of a new chiral invariant
operator leads to the opening of the new absorptive cuts
in addition to those  channels we are in principle interested in.
%%
%%For instance, at large $N_C$ one needs the $c_m\bra S \chi_+\ket $
%%to describe the decay of the $\bar{q}q$ scalar density into one
%%scalar resonance, but at the same time it also generates the  decay into  $S\pi$
%%(although other operators like $\lambda_1^{SP}$ become also relevant
%%in this decay).
%%
%%In the same way,
%%
For instance, the $c_m\bra S \chi_+\ket $ rules the decay into one
scalar resonance and also contributes to the $S$-meson exchange in
the $\pi\pi$ channel. But at the same time it also induces  the
decay into  $S\pi$ (though other operators like $\lambda_1^{SP}$ are
also relevant).
Thus, the $\mL_{RR'}$ terms were used in our
calculation to improve the description of the $R\pi$ channels, which
were incompletely described by the linear lagrangian $\mL_R$~\cite{Ecker1}.
The price to pay was that  new $RR'$ channels with two intermediate
resonances showed up in our NLO computation of the correlator.
Although the impact of these higher thresholds is suppressed at low
energies if one chooses a convenient renormalization scheme~\cite{L10,Rosell-thesis},
their impact in the high-energy
matching and OPE constraints is a priori  non-trivial. In this paper
we find that, indeed,  the most relevant information in order to
extract the low energy chiral couplings seems to be  provided by the
lightest cuts.   On the other hand, one realizes that the values of
the couplings differ from
those in the full large--$N_C$ theory~\cite{Golterman-L8}
and that the description of
the heaviest absorptive channels may be very
distorted~\cite{Masjuan-Pade}. Indeed, we obtain the resonance
couplings $\lambda_1^{SP}=-0.22\pm 0.08$, $\lambda_1^{PV}=0.14\pm
0.07$ and $|\lambda_1^{SA}|=0.16\pm 0.14$.  Even though these
numbers have the right signs and order of the magnitude as the
theoretical expectations $\lambda_1^{SP}=-\frac{d_m}{c_m}=\frac{c_d
-2 c_m}{2 d_m}\sim -0.7$, $\lambda_1^{PV}=\frac{G_V}{2\sqrt{2}
d_m}\sim 0.7$ and $\lambda_1^{SA}=0$ (in our analysis, for convention,
we have took $c_m$, $d_m$ and $G_V$ as positive),  their
values are still far from being accurate determinations of these
parameters.

\subsection{Impact of the $RR'$ channels}

In this section we will make a digression on
the importance of the $RR'$ intermediate cuts
that are opened after including the $\mL_{RR'}$ operators in the
LO action.  We will remove by hand the contributions with  two--resonance cuts.
Although this procedure is not well justified from the QFT
point of view, we will perform this exercise in order
make a rough comparison with the previous dispersive calculation of the octet
$SS-PP$ correlator~\cite{L8}. The $RR'$ channels were neglected there,
as their contribution in the dispersive integral
was suppressed at low energies by inverse powers of $(M_R+M_{R'})^2$.

Thus, we redid the calculation and removed by hand the diagrams with
two--resonance cuts.  This expression was then matched to the OPE
at short distances,  producing finally the low--energy constants,
\begin{equation}
L_8(\mu_0)\,=\, (0.1\, \pm\, 0.7)\, \cdot\, 10^{-3}\, , \qquad\qquad
C_{38}(\mu_0)\, =\, (-3\,\pm\, 9)\,\cdot\, 10^{-6}\, ,
\end{equation}
where we used the same inputs as in the previous subsection.
The errors are now found to be larger and, though compatible with
our final result~(\ref{eq.final-LEC}),
the elimination of the $RR'$ cuts decreases slightly  the range
for the LEC determinations,
approaching them to the lower values preferred by recent $\cO(p^6)$
analysis~\cite{new-fit10}
and lattice simulations~\cite{MILC}.
However, discarding these heavier channels from the one-loop computation
in this way does not seem very sound from the theoretical point of view
and it is shown here just as an exercise.

\section{Conclusions}
\label{sec.conclu}

In this paper, we have performed the one loop QFT calculation of
the two-point $SS-PP$ correlator within R$\chi$T.
We started with Ecker {\it et al.}'s lagrangian~\cite{Ecker1}, containing only operators with
at most one resonance field,  and
renormalized step by step all the relevant vertex-functions
and propagators. Then we
imposed OPE constraints on the full one-loop correlator,
not on separate individual
channels as it was performed in a previous NLO calculation~\cite{L8}.
Likewise, no short-distance constraint from other observables~\cite{PI:08}
was used in the present article.

After fixing part of our \rcht\ couplings through these high-energy conditions,
we expanded  our result at low energies.
Due to the chiral invariant structure of R$\chi$T, we were able
to match the chiral logarithms  and found predictions for
the $\chi$PT coupling constants $L_8(\mu)$ and $C_{38}(\mu)$.
The large discrepancy of these first  numerical determinations
with respect to the numbers found in the literature indicated that the
simple Lagrangian $\mL_R$ (with operators with  at most one resonance
field~\cite{Ecker1})  pointed out the need for a more complicated
structure of the \rcht\ action.
The $\mL_R$ terms could  not fully describe the dynamics of
all the two-meson intermediate channels:
just the  $\pi\pi$ channel description was adequately provided
by the operators with at most one resonance field;
all other channels ($V\pi$, $S\pi$\dots) did not have
the right short-distance behavior.
Thus, beyond any numerical  discrepancy in the LECs, the absence of operators with two an three resonance fields
produces  a severe theoretical issue at high energies~\cite{L9}.

In order to arrange the $R\pi$ cuts with one resonance and one Goldstone
we add all the operators $\mL_{RR'}$ with two resonance fields
relevant for the $SS-PP$ correlator to the leading \rcht\ lagrangian.
These are the $\lambda_1^{SP}$, $\lambda_1^{PV}$ and $\lambda_1^{SA}$
terms given in Eq.~(\ref{eq.lambda-lagr}).
The introduction of these operators produce a dramatic improvement.
When only one of them is added to the action, the LEC predictions
move in the right direction, i.e., towards the range of values found in
previous studies.
After considering all the three $\mL_{RR'}$ operators,
we obtain the final  values for $\mu_0=770$~MeV,
\begin{equation}
L_8(\mu_0)\,\, =\,\, (1.0\, \pm\,  0.4)\, \cdot\,  10^{-3}\, , \qquad\qquad
C_{38}(\mu_0) \,\, =\,\,  (8\, \pm\,  5)\, \cdot\,  10^{-6}\, ,
\end{equation}
in reasonable agreement with the values obtained through other
approaches~\cite{SS-PP-Bijnens,fit10,new-fit10,L8,DSE,MILC}.
We want to remark, that this result is progressively approached
as more and more complicated operators are added to the hadronic action.
The terms of the lagrangian that rule the lightest channels
result crucial and, thus, those determining heavier cuts  not included
in the analysis are expected to produce little influence.

The essential difference with the previous dispersive calculation of the
$SS-PP$ correlator at NLO~\cite{L8} is the presence of  $RR'$ cuts in
the present work. These intermediate channels automatically show up
at the very moment we place the $\mL_{RR'}$
operators in the \rcht\ action.  Although it is possible to demonstrate
that the contribution from these heavy $RR'$ cuts is suppressed
at low energies~\cite{L10,Rosell-thesis},
their impact in high-energy conditions such as the NLO Weinber sum-rules
is pretty non-trivial.  The difference between the present article and
Ref.~\cite{L8} could be taken as a crude estimate of the impact
of neglecting those higher channels.

In addition to the estimation of LECs,   we also discussed
some general issues about renormalization schemes within
R$\chi$T.
The use of the running $\wms$ masses $M_R(\mu)$ was not very convenient
as their meaning changed as one added new operators to the \rcht\ action.
Thus, they were reexpressed in terms of pole masses $\hat{M}_R$.
Likewise, we found that, with respect to the large--$N_C$ WSR,
the NLO Weinberg sum-rules~(\ref{eq.NLO-WSR})
led to large uncertainties and variations for the values
of $c_m$ and $d_m$ derived from them in the $\wms$--scheme.
A more convenient subtraction scheme was found to minimize these
uncertainties that stemmed from the high-energy matching
whereas, on the other hand, it was found to leave the low energy
prediction~(\ref{eq.L8-linf}) unchanged  (except for the improved accuracy
in the resonance coupling determination from short-distance constraints).

\section*{Acknowledgement}

We would like to thank K. Kampf, J. Novotny, S. Peris and I. Rosell
for useful discussions and valuable comments on the manuscript.
This work is supported in part
by the Center for Particle Physics (Project no. LC 527), GAUK
(Project no.6908; 114-10/258002),
CICYT-FEDER-FPA2008-01430,
SGR2005-00916, SGR2009-894, the Spanish Consolider-Ingenio 2010 Program CPAN
(CSD2007-00042), the Juan de la Cierva Program and the EU Contract
No. MRTN-CT-2006-035482, "FLAVIAnet". J. T. is also supported by the
U.S. Department of State (International Fulbright S\&T award).

\appendix

\section{Running of the renormalized parameters with $\mL_G+\mL_R$}

When only operators with at most one resonance fields are considered
in the \rcht\ action~\cite{Ecker1},  one finds
before performing the meson field redefinition the running,
\begin{eqnarray}
\Frac{\partial\widetilde{L}_8}{\partial\ln\mu^2} &=&
\frac{3}{512\pi^2F^2}\left[16(c_m^2-d_m^2)-F^2-16c_dc_m +4c_d^2 +2G_V^2\right] \, ,
\nonumber\\
\Frac{\partial M_S^2}{\partial\ln\mu^2}  &=&\Frac{\partial M_P^2}{\partial\ln\mu^2}
\,=\,\Frac{\partial c_m}{\partial\ln\mu^2}\,=\,\Frac{\partial d_m}{\partial\ln\mu^2}
\,=\, 0\, ,
\nonumber\\
\Frac{\partial X_S}{\partial\ln\mu^2}   &=&  - \frac{3c_d^2}{16\pi^2F^4}\,,
\qquad\qquad \qquad
\Frac{\partial X_P}{\partial\ln\mu^2} = 0\,,
\nonumber\\
\Frac{\partial \lambda^S_{18}}{\partial\ln\mu^2} &=& \frac{3c_d}{64\pi^2F^2}\,,
\qquad\qquad\qquad
\Frac{\partial \lambda^P_{13}}{\partial\ln\mu^2} = 0
\,.
\end{eqnarray}

After the renormalization one may then consider a convenient field
redefinition that removes precisely the renormalized $X_{S,P}$,
$\lambda_{13}^P$ and $\lambda_{18}^S$.  They (and their running)
seem to disappear from the theory although their information is
actually encoded in the renormalized effective couplings that remain
in the action. Their running turns out to be then
\begin{eqnarray}
\Frac{\partial \widetilde{L}^{\rm eff}_8}{\partial\ln\mu^2}
&=&  \frac{3}{512\pi^2F^2}
\left[16 \,c_m^2- 16\,d_m^2 -F^2-8\, c_d c_m +2\,c_d^2+2\,G_V^2 -16\, c_d^2c_m^2\right]\,,
\nonumber\\
\Frac{\partial M_S^{{\rm eff}\,\, 2}}{\partial\ln\mu^2} &=&
\frac{3c_d^2M_S^4}{16\pi^2F^4} \,,
\qquad\qquad\qquad \qquad\qquad\qquad
\Frac{\partial M_P^{{\rm eff}\,\, 2}}{\partial\ln\mu^2} = 0\,,
\nonumber\\
\Frac{\partial c_m^{\rm eff}}{\partial\ln\mu^2}   &=&
\frac{3c_dM_S^2}{64\pi^2F^4}(4c_dc_m-F^2)\,,
\qquad\qquad\qquad
\Frac{\partial d_m^{\rm eff}}{\partial\ln\mu^2} =0\,.
\end{eqnarray}

\section{On-shell scheme for $c_m$ and $d_m$}

This would be a continuation of the pole-mass scheme. In addition to
this, the renormalized on-shell couplings $\hat{c}_m$ and
$\hat{d}_m$ are prescribed, respectively, by the real part of the
residue of the correlator at the scalar and the pseudoscalar
resonance poles~\cite{L8,L10}. This was the scheme considered in the
dispersive approach from Refs.~\cite{L8,L10}. The shift $\Delta
\kappa$ with respect to the $\wms$--subtraction prescription is
given up to NLO in $1/N_C$ by
\begin{eqnarray}
2 \, c_m \Delta c_m &=& c_m^2 - \hat{c}_m^{2}\,=\,
\Frac{c_m}{2 B_0}\,\mbox{Re}\Phi_{sS}^r(M_S^2)^{1\ell}\,\,
-\,\, c_m^2\, \mbox{Re}\Sigma_S^{r\,\,'}(M_S^2)^{1\ell}\, ,
\nn\\
2 \,  d_m \Delta d_m &=& d_m^2 - \hat{d}_m^{2}\,=\,
\Frac{d_m}{2 B_0}\,\mbox{Re}\Phi_{pP}^r(M_P^2)^{1\ell}\,\,
-\,\, d_m^2\, \mbox{Re}\Sigma_P^{r\,\,'}(M_P^2)^{1\ell}\, .
\end{eqnarray}

In the case where only $\mL_R$ interactions are considered, one has
\begin{eqnarray}
2 \,  c_m \Delta c_m &=&
\Frac{4 \, c_d c_m}{F^2}\,\Frac{ 3 \, M_S^2}{128\pi^2}
\left[ -1 +\left(1-\Frac{4 \, c_d c_m}{F^2}\right) \ln\Frac{M_S^2}{\mu^2}\right]
\,,
\nn\\
2  \, d_m \Delta d_m &=& 0\, .
\end{eqnarray}

\section{Feynman integrals}
\label{app.Feynman}

The scalar integrals are
\begin{eqnarray}
A_0(M^2) &=& \int\frac{dk^d}{i(2\pi)^d} \frac{1}{k^2-M^2+i\epsilon}\nonumber,\\
B_0(p^2,M_a^2,M_b^2) &=& \int
\frac{dk^d}{i(2\pi)^d}\frac{1}{(k^2-M_a^2+i\epsilon)[(p-k)^2-M_b^2+i\epsilon]}
\end{eqnarray}
Using the formula in \cite{L9} we use the following expansions
\begin{eqnarray}
A_0(M^2) &=& \frac{-M^2}{16\pi^2}\Bigg\{\lambda_\infty + \ln\frac{M^2}{\mu^2}\Bigg\}\nonumber,\\
B_0(p^2,0,0) &=& -\frac{1}{16\pi^2}\Bigg\{ \lambda_\infty -1 +\ln\left(\frac{-p^2}{\mu^2}\right)\Bigg\}\nonumber,\\
B_0(p^2,0,M^2) &=& -\frac{1}{16\pi^2}\Bigg\{ \lambda_\infty +\ln\frac{M^2}{\mu^2} -1 + \left(1-\frac{M^2}{p^2}\right)
\ln\left(1-\frac{p^2}{M^2}\right)\Bigg\}\nonumber,\\
B_0(p^2,M^2,M^2) &=& -\frac{1}{16\pi^2}
\Bigg\{ \lambda_\infty +\ln\frac{M^2}{\mu^2} -1 +\sigma_M\ln\left(\frac{\sigma_M+1}{\sigma_M-1}\right)\Bigg\}
\\
\overline{J}(p^2,M_a^2,M_b^2) &=&
\frac{1}{32\pi^2}
\left\{
2 +\left[ \Frac{M_a^2-M_b^2}{p^2}\, -\, \Frac{M_a^2+M_b^2}{M_a^2-M_b^2}
\right]\ln\Frac{M_b^2}{M_a^2}
\right.
\nn\\
&&
\left.
\, -\,\Frac{\lambda^{1/2}(p^2,M_a^2,M_b^2) }{p^2}
\, \ln\left(
\Frac{\left[p^2+\lambda^{1/2}(p^2,M_a^2,M_b^2)\right]^2-(M_a^2-M_b^2)^2}{
\left[p^2-\lambda^{1/2}(p^2,M_a^2,M_b^2)\right]^2-(M_a^2-M_b^2)^2}
\right)\right\}
\, ,  \nn
\end{eqnarray}
where $\sigma_M = \sqrt{1-4M^2/p^2}$ and
$\lambda(x,y,z)=(x-y-z)^2-4yz$.

\section{Useful expansions}

Using expansions for $x\rightarrow \infty$
\begin{eqnarray*}
\phi(x) &=& \ln(-x) - 1 - \frac{3\ln(-x)}{x} +\frac{3}{2x}
+\frac{3\ln(-x)}{x^2} +\frac{3}{2x^2} + \dots,\\
\psi(x) &=& -\ln(-x) + \frac{2\ln(-x)}{x} - \frac{\ln(-x)}{x^2}
-\frac{3}{2x^2} + \dots,\\
(1-\frac{1}{x})\ln(1-x) &=& \ln(-x) -\frac{\ln(-x)}{x} -\frac{1}{x}
+\frac{1}{2x^2}+\dots
\end{eqnarray*}

\end{document}